\documentclass[a4paper,fleqn,usenatbib]{mnras}
\pdfoutput=1

\usepackage{newtxtext,newtxmath}

\usepackage[T1]{fontenc}
\usepackage{ae,aecompl}

\usepackage{graphicx}	
\usepackage{amsmath}	
\usepackage{amssymb}	
\usepackage{xspace}     
\usepackage{comment}
\usepackage{booktabs}

\newcommand{\msun}{{\mathrm M}_{\odot}}
\newcommand{\lsun}{{\mathrm L}_{\odot}}
\newcommand{\rsun}{{\mathrm R}_{\odot}}

\newcommand{\kms}{\mathrm{km}\,\mathrm{s}^{-1}}
\newcommand{\yr}{\mathrm{yr}}
\newcommand{\myr}{\mathrm{Myr}}
\newcommand{\gauss}{\mathrm{G}}
\newcommand{\oned}{1D\@\xspace}

\newcommand{\threed}{3D\@\xspace}
\newcommand{\alfven}{Alfv\'en\@\xspace}
\newcommand{\tausco}{$\tau$~Sco\@\xspace}
\newcommand{\etacar}{$\eta$~Car\@\xspace}

\newcommand{\mesa}{\mbox{\textsc{Mesa}}\xspace}
\newcommand{\arepo}{\mbox{\textsc{Arepo}}\xspace}

\newcommand*{\eg}{e.g.\@\xspace}
\newcommand*{\ie}{i.e.\@\xspace}
\newcommand*{\cf}{cf.\@\xspace}
\newcommand{\tnu}{\tilde\nu}

\title[Evolution of a magnetic massive merger product]{Long-term evolution of a magnetic massive merger product}

\author[Fabian~R.N.~Schneider~et~al.]{
F.R.N. Schneider$^{1,2,3}$\thanks{E-mail: fabian.schneider@uni-heidelberg.de},
S.T. Ohlmann$^{4}$,
Ph. Podsiadlowski$^{3}$,
F.K. R{\"o}pke$^{2,5}$,\newauthor 
\;S.A. Balbus$^{3}$, and
R. Pakmor$^{6}$
\\
$^{1}$Zentrum f\"{u}r Astronomie der Universit\"{a}t Heidelberg, Astronomisches Rechen-Institut, M\"{o}nchhofstr. 12-14, 69120 Heidelberg, Germany\\
$^{2}$Heidelberger Institut f\"{u}r Theoretische Studien, Schloss-Wolfsbrunnenweg 35, 69118 Heidelberg, Germany\\
$^{3}$Department of Physics, University of Oxford, Denys Wilkinson Building, Keble Road, Oxford OX1~3RH, United Kingdom\\
$^{4}$Max Planck Computing and Data Facility, Gie\ss{}enbachstr.\ 2, 85748 Garching, Germany\\
$^{5}$Zentrum f\"{u}r Astronomie der Universit\"{a}t Heidelberg, Institut f\"{u}r Theoretische Astrophysik, Philosophenweg 12, 69120 Heidelberg, Germany\\
$^{6}$Max-Planck-Institut f{\"u}r Astrophysik, Karl-Schwarzschild-Str.\ 1, 85748 Garching, Germany\\
}

\date{Accepted 2020 May 7. Received 2020 April 28; in original form 2020 March 2}

\pubyear{2020}

\begin{document}
\label{firstpage}
\pagerange{\pageref{firstpage}--\pageref{lastpage}}
\maketitle

\begin{abstract}
About 10\% of stars more massive than ${\approx}\,1.5\,\msun$ have strong, large-scale surface magnetic fields and are being discussed as progenitors of highly-magnetic white dwarfs and magnetars. The origin of these fields remains uncertain. Recent \threed magnetohydrodynamical simulations have shown that strong magnetic fields can be generated in the merger of two massive stars. Here, we follow the long-term evolution of such a \threed merger product in a \oned stellar evolution code. During a thermal relaxation phase after the coalescence, the merger product reaches critical surface rotation, sheds mass and then spins down primarily because of internal mass readjustments. The spin of the merger product after thermal relaxation is mainly set by the co-evolution of the star-torus structure left after coalescence. This evolution is still uncertain, so we also consider magnetic braking and other angular-momentum-gain and -loss mechanisms that may influence the final spin of the merged star. Because of core compression and mixing of carbon and nitrogen in the merger, enhanced nuclear burning drives a transient convective core that greatly contributes to the rejuvenation of the star. Once the merger product relaxed back to the main sequence, it continues its evolution similar to that of a genuine single star of comparable mass. It is a slow rotator that matches the magnetic blue straggler \tausco. Our results show that merging is a promising mechanism to explain some magnetic massive stars and it may also be key to understand the origin of the strong magnetic fields of highly-magnetic white dwarfs and magnetars.
\end{abstract}




\section{Introduction}\label{sec:intro}

Since the discovery of a strong, large-scale surface magnetic field in the A-type star 78~Vir by \citet{1947PASP...59..112B}, large surveys have revealed that about 10\% of all intermediate-mass and high-mass stars (${\gtrsim}\,1.5\,\msun$; referred to as ``massive'' stars hereafter) have such strong magnetic fields \citep[see \eg][]{2009ARA&A..47..333D,2015A&A...582A..45F,2017MNRAS.465.2432G}. It has been suggested that these fields are inherited from a magnetised molecular cloud or are generated in-situ by a dynamo \citep[\eg][]{1999stma.book.....M,2001ASPC..248....3M,2001ASPC..248..305M}. Also, mergers of stars have been put forward as a mechanism to form strong magnetic fields \citep[\eg][]{2005MNRAS.356..615F,2005MNRAS.356.1576W,2009MNRAS.400L..71F,2012ARA&A..50..107L,2014IAUS..302....1L,2014MNRAS.437..675W}. By means of \threed magnetohydrodynamical (MHD) simulations of the merger of two massive main-sequence (MS) stars, \citet{2019Natur.574..211S} could show that this indeed appears to be a viable channel.

The merger hypothesis can naturally explain why only a rather small fraction of massive stars have strong surface magnetic fields and also why there is a dearth of magnetic stars in close binaries \citep{2002A&A...394..151C,2015IAUS..307..330A}. Stars can merge during their pre-MS and MS phases. In the latter case, the magnetic merger product rejuvenates and may stand out as a blue straggler in a star cluster or association. \citet{2016MNRAS.457.2355S} identified two such candidates, the magnetic B-type star HR~2949, which appears anomalously young compared to its visual binary companion HR~2948, and the former B0 spectral standard star \tausco, a blue straggler in the ${\approx}\,11\,\myr$ old Upper Scorpius association \citep{2012ApJ...746..154P}. The age discrepancies found in these two stars are consistent with being due to rejuvenation in a merger \citep{2016MNRAS.457.2355S}.

Motivated by this, \citet{2019Natur.574..211S} \citep[see also][]{Ohlmann+2019} carried out \threed MHD merger simulations of MS stars of $9\,\msun$ and $8\,\msun$ to obtain a merger product that closely resembles \tausco, in particular with respect to its apparently young age. To this end, the merger was initiated at an age of $9\,\myr$. This means that the merging stars were relatively unevolved and had finished only 31\% and 25\% of their MS evolution, respectively. The simulations were carried out with the moving-mesh code \arepo \citep{2010MNRAS.401..791S,2011MNRAS.418.1392P,2013MNRAS.432..176P} and showed the rapid amplification of a magnetic field to maximum field strengths of $10^7\text{--}10^8\,\mathrm{G}$, rendering merging a promising model to explain the enigmatic magnetic massive stars. In the amplification of the magnetic field, the magneto-rotationally instability \citep[MRI;][]{1991ApJ...376..214B,1991ApJ...376..223H} seemed to play a key role. 

Magnetic stars have been suggested as progenitors of highly magnetic white dwarfs and magnetars \citep[\eg][]{2005MNRAS.356..615F,2005MNRAS.356.1576W,2004MNRAS.355L..13T,2014MNRAS.437..675W}. Indeed, the magnetic flux in the core of the merger remnant produced in the simulations of \citet{2019Natur.574..211S} is sufficiently high to explain the strong magnetic fields of $10^{13}\text{--}10^{15}\,\gauss$ inferred to exist on the surfaces of magnetars \citep{2002ApJ...574L..51I,2014ApJS..212....6O}.

In this work, we start out from the results of the \threed merger simulation of \citet{2019Natur.574..211S} and \citet{Ohlmann+2019} and model the long-term evolution of the strongly magnetised merger remnant with the \oned stellar evolution code \mesa \citep{2011ApJS..192....3P,2013ApJS..208....4P,2015ApJS..220...15P,2018ApJS..234...34P,2019ApJS..243...10P}. In Sect.~\ref{sec:methods}, we describe the physical setup of our \oned merger models and then, in Sect.~\ref{sec:merger-evolution}, we describe their evolution through a thermal relaxation phase up to core helium exhaustion. Key aspects of the evolution are discussed in Sect.~\ref{sec:discussion} and we conclude in Sect.~\ref{sec:conclusions}. Some features of the default merger model presented here have been described in \citet{2019Natur.574..211S}.

\section{Methods}\label{sec:methods}

Modelling the long-term evolution of a magnetised merger remnant from \threed simulations in a \oned stellar evolution code requires a few approximations and extensions to the usual physics captured by \oned models. This mainly relates to the amplified magnetic field and its influence on the evolution of the star, here especially the stellar spin. A brief description of the main \mesa setup is given in Sect.~\ref{sec:mesa-setup} while the extensions are described in Sects.~\ref{sec:b-field}--\ref{sec:mass-accretion}. Because it is not possible to self-consistently include and evolve a \threed magnetic field in a \oned stellar evolution code, we consider that there exists a large-scale, dipolar magnetic field (Sect.~\ref{sec:b-field}) that contributes to the transport of angular momentum in the stellar interior (Sect.~\ref{sec:ang-mom-transport}) and to enhanced angular-momentum loss from the star through the coupling of the magnetic field to the stellar wind (magnetic braking; Sect.~\ref{sec:magnetic-braking}). Except for these two aspects, the magnetic field does not influence the evolution of the merger product in our models. There are situations where the merger product may accrete from a disk (Sect.~\ref{sec:mass-accretion}) and the transition from the \threed to \oned models needs careful consideration (Sect.~\ref{sec:3d-to-1d}).

\subsection{Main \mesa model setup}\label{sec:mesa-setup}

We use \mesa version 9793. The models assume the solar composition of \citet{2009ARA&A..47..481A} with a metallicity $Z=0.0142$ and initial helium mass fraction $Y=0.2703$ with corresponding \mesa opacity tables as outlined in the \mesa instrument papers \citep[low temperature opacities are from][]{2005ApJ...623..585F}. Convection is treated within mixing-length theory \citep{1965ApJ...142..841H} and we use the Ledoux criterion for convective boundaries, a mixing-length parameter of $\alpha_\mathrm{MLT}=1.8$ and a semi-convection efficiency of $\alpha_\mathrm{sm}=1.0$ (we do not use \mesa's MLT++ scheme). Exponential convective core-overshooting \citep[\eg][]{2000A&A...360..952H} is employed with an overshooting parameter of $f_\mathrm{over}=0.0194$, which results in a MS width that is reminiscent of the empirical MS width of Galactic ${\approx}\,10\,\msun$ stars found by \citep{2014A&A...570L..13C}. Shellular rotation is assumed \citep[\eg][]{1970A&A.....5..155K,1976ApJ...210..184E,1997A&A...321..465M,2000ApJ...528..368H} and rotational mixing is included via the Solberg--H{\o}iland instability, the Goldreich--Schubert--Fricke instability, Eddington--Sweet circulation, the secular-shear instability, and the dynamical shear instability as detailed in \citet{2000ApJ...528..368H}. The rotational-mixing efficiency is set to $f_\mathrm{mix}=0.0333$ and the inhibiting chemical gradient is scaled with a factor of $f_\mu=0.1$. The Spruit--Taylor dynamo \citep{2002A&A...381..923S} is not included. \mesa's 'Dutch' wind mass-loss prescription (scaling factor of 1) is used, \ie \citet{2000A&A...362..295V,2001A&A...369..574V} for hot stars, \citet{1988A&AS...72..259D} for cool stars and \citet{2000A&A...360..227N} for Wolf--Rayet stars. For stars approaching critical rotation, we apply an implicit method that removes as much mass as is needed to keep stars rotating below a certain fraction (here 99\%) of critical rotation \citep{2015ApJS..220...15P}. Nuclear burning is via \mesa's 'basic.net' reaction network that includes the most important reactions for hydrogen und helium burning. Consequently, we stop the evolution of our models upon core helium exhaustion (when the central helium mass fraction falls below $10^{-4}$).

\subsection{Assumed large-scale magnetic field}\label{sec:b-field}

The \threed MHD simulation has shown that a strong magnetic field is created in the merger. As yet, we cannot predict how this field evolves after the merger and how it would interact with other magnetic fields produced in-situ by \eg the Spruit--Taylor dynamo \citep{2002A&A...381..923S,2005ApJ...626..350H}, the MRI \citep{2015ApJ...799...85W}, or other dynamo processes \citep[\eg a radiative $\alpha$--$\Omega$ dynamo,][]{2012MNRAS.424.2358P,2018MNRAS.477.2298Q}. Because of this, we here consider a static magnetic field that will contribute to angular-momentum transport in the stellar interior, but we assume that there are no other dynamo-generated fields. For the magnetic field geometry and strength in the \oned stellar evolution models, we assume that a static, large-scale, dipole magnetic field permeates the star such that the absolute magnetic field strength $B$ as a function of radius $r$ follows
\begin{equation}
B(r) = \mu_\mathrm{B} r^{-3},
\label{eq:dipole-b-field}
\end{equation}
where $\mu_\mathrm{B}=B_*R_*^3$ is the magnetic dipole moment, $B_*$ the surface magnetic field and $R_*$ the radius of the star. For a dipole magnetic field, the ratio of the field strength at the pole ($B_\mathrm{p}$) and the equator ($B_\mathrm{eq}$) is $B_\mathrm{p}/B_\mathrm{eq}=2$. The dipole field diverges for $r\rightarrow 0$ and we therefore limit its field strength to $10^9\,\gauss$.

We assume that the magnetic field cannot penetrate convective regions if the convective energy density is larger than the magnetic energy density, \ie $1/2 \rho v_{\mathrm{conv}}^{2} > B^{2}/8\pi$ (known as convective expulsion, see \eg \citealt{Zeldovich1957}, \citealt{1963ApJ...138..552P}, \citealt{1966RSPSA.293..310W}, and the review by \citealt{2017RSOS....460271B}). Also, some of the aforementioned dynamo-generated magnetic fields have the tendency to be suppressed in convective regions \citep[\eg][]{2018MNRAS.477.2298Q}. Here, $\rho$ is the gas density and $v_{\mathrm{conv}}$ the velocity of convective eddies as predicted by mixing length theory. Essentially, this means that the magnetic field can only contribute to the angular-momentum transport in radiative regions and does not provide an efficient coupling of, \eg, the convective core and the radiative envelope. 

A dynamo process potentially operating in the convective core may be influenced by a large-scale magnetic field in the radiative outer regions of a star \citep{2009ApJ...705.1000F}. Such interactions and the possibility that the large-scale magnetic field may damp the convective motions in or near the stellar core \citep[\eg limit convective core overshooting or even decrease the size of the convective core; see][and references therein]{2015A&A...584A..54P} are not accounted for in our model.

The evolution of the \oned merger product depends on the assumed strength of the magnetic field because of the interior angular-momentum transport via the magnetic field and magnetic braking. In the following, we consider three cases:
\begin{enumerate}
\item A weak magnetic field with $\mu_\mathrm{B}=2\times 10^{34}\,\mathrm{G}\,\mathrm{cm}^3$ that corresponds to a polar magnetic field strength of $1\,\mathrm{G}$ for a $4\,\rsun$ star, similar to Vega-like, sub-Gauss magnetic fields that might be ubiquitous in stars with a radiative envelope \citep[\eg][]{2009A&A...500L..41L,2011A&A...532L..13P}.
\item A magnetic dipole moment of $\mu_\mathrm{B}=2\times 10^{37}\,\mathrm{G}\,\mathrm{cm}^3$ that is reminiscent of that of \tausco \citep[polar field strength of ${\approx}\,500\,\mathrm{G}$ for a ${\approx}\,5\,\rsun$ star; \eg][]{2006MNRAS.370..629D}. This is our default model.
\item A strong magnetic field with $\mu_\mathrm{B}=1\times 10^{40}\,\mathrm{G}\,\mathrm{cm}^3$ that can be compared to the magnetic field found immediately after the merger of the two MS stars considered in this work \citep{2019Natur.574..211S,Ohlmann+2019}.
\end{enumerate}

\subsection{Angular-momentum transport in the stellar interior through a large-scale magnetic field}\label{sec:ang-mom-transport}

We follow \citet{2019Natur.574..211S} and treat the angular-momentum transport by the magnetic field through the stellar interior as a diffusive process with an effective diffusion coefficient (effective magnetic viscosity $\tnu_\mathrm{eff}$) of
\begin{equation}
\tnu_{\mathrm{eff}} = \frac{3I}{8\pi r^{4}\rho}v_{\mathrm{A}}.
\label{eq:nu-eff}
\end{equation}
In this equation, $I$ is the moment of inertia of a shell in the stellar interior, $r$ its radial coordinate, $\rho$ its density and $v_\mathrm{A}=B/\sqrt{4\pi\rho}$ the local \alfven velocity. We assume that `shells' in Eq.~(\ref{eq:nu-eff}) have a thickness of 20\% of the local pressure scale height $H_\mathrm{P}$; this is to ensure that $\tnu_{\mathrm{eff}}$ does not depend on numerical resolution. Whenever we refer to viscosity in this paper, we do not mean the microscopic viscosity (which is negligible in our situation), but we rather have an effective viscosity in mind that can act on larger scales and is caused by an enhanced turbulent transport.

For thin shells of mass $\Delta m$, the moment of inertia is $I=2/3 \Delta m r^{2}$. The effective magnetic viscosity is then approximately $\tnu_{\mathrm{eff}} = \Delta r v_{\mathrm{A}}$ with $\Delta r=0.2 H_\mathrm{P}$. This is similar to the general form of a diffusion coefficient, $D \propto l v$, for a diffusion process over a length scale $l$ with characteristic velocity $v$. We modulate the effective turbulent viscosity with a factor $f_{\mathrm{A}}$ that is thought to adjust the timescale over which solid-body rotation is achieved in neighbouring shells. We set $f_{\mathrm{A}}=0.5$ in our calculations. Taken together with our choice of $f_{\mathrm{A}}$, the effective magnetic viscosity is equivalent to $\tnu_\mathrm{eff} \approx 0.1 H_\mathrm{P} v_\mathrm{A}$.

\subsection{Magnetic braking}\label{sec:magnetic-braking}

For a rotating star with a large-scale surface magnetic field, the stellar wind may bend magnetic-field lines which leads to Poynting stresses and hence a torque on the stellar surface \citep[\eg][]{1967ApJ...148..217W,1987MNRAS.226...57M}. The torque from such magnetic braking is usually written in the form
\begin{equation}
\frac{\mathrm{d}J_{\mathrm{mb}}}{\mathrm{d}t} = \frac{2}{3} \dot{M} \Omega_{*} R_{\mathrm{A}}^{2},
\label{eq:magnetic-braking}
\end{equation}
where $\dot{M}$ is the stellar wind mass-loss rate, $\Omega_{*}$ is the stellar surface angular velocity, $R_{\mathrm{A}}$ is the \alfven radius and the factor $2/3$ accounts for the moment of inertia of a thin spherical shell. This form of the torque resembles the idea that a surface magnetic field forces the stellar wind into solid-body rotation with the star out to the \alfven radius such that the wind removes the specific angular momentum from the \alfven surface and not from the stellar surface.

Within this formalism, MHD simulations of magnetic braking in hot, massive stars \citep{2009MNRAS.392.1022U} have shown that
\begin{equation}
\frac{R_{\mathrm{A}}}{R_{*}} \approx 0.29+\left(\eta_{*}+0.25\right)^{1/4}.
\label{eq:alfven-radius-mb}
\end{equation}
Here, $R_{*}$ is the stellar radius and $\eta_{*}$ is the so-called wind magnetic confinement parameter defined by
\begin{equation}
\eta_{*}=\frac{B_{\mathrm{eq}}^{2}R_{*}^{2}}{\dot{M}v_{\infty}}
\end{equation}
with the terminal wind velocity $v_{\infty}$. 

We take the observed terminal wind velocities of stars from spectral types O to F from \citet{1995ApJ...455..269L}, which, \eg, are also used in \citet{2000A&A...362..295V} to derive the mass-loss rates of hot stars across the bi-stability jump,
\begin{equation}
\frac{v_{\infty}}{v_{\mathrm{esc}}}=\begin{cases}
0.7 & \text{for }\log T_{\mathrm{eff}}\leq4.0,\\
1.3 & \text{for }4.0<\log T_{\mathrm{eff}}\leq4.32,\\
2.6 & \text{for }4.32<\log T_{\mathrm{eff}}.
\end{cases}
\end{equation}
Here, the escape velocity is defined as
\begin{equation}
v_{\mathrm{esc}}=\sqrt{\frac{2GM(1-\Gamma_{\mathrm{es}})}{R_{*}}}
\end{equation}
with $\Gamma_{\mathrm{es}}$ the electron-scattering Eddington factor and $M$ the stellar mass.

The resulting spin-down timescale through magnetic braking then is \citep[Eq.~21 of][]{2009MNRAS.392.1022U},
\begin{equation}
\tau_\mathrm{spin-down} = \frac{3/2 r_\mathrm{g}^2}{\left[0.29+\left(\eta_{*}+0.25\right)^{1/4}\right]^2} \tau_{\dot M},
\label{eq:spin-down-timescale}
\end{equation}
where $r_\mathrm{g}$ is the (relative) radius of gyration ($r_\mathrm{g}^2 = I/MR^2$ is also known as the moment of inertia factor) and $\tau_{\dot M}=M/{\dot M}$ the mass-loss timescale.

\subsection{Mass accretion from a disk}\label{sec:mass-accretion}

In this work, we encounter situations where mass is accreted from a disk onto a central star. In these situations, we assume that the disk can be described as a geometrically thin and optically thick Keplerian disk. From angular-momentum conservation in such a disk, the radial drift velocity is \citep[\eg][]{1981ARA&A..19..137P},
\begin{equation}
v_{\mathrm{r}}=-\frac{3}{\Sigma r^{1/2}}\frac{\partial}{\partial r}\left[\tnu\Sigma r^{1/2}\right],
\label{eq:ang-mom-cons-disk}
\end{equation}
where $\Sigma$ is the surface mass density, $r$ the radius and $\tnu$ the effective turbulent viscosity of the disk. We assume the $\alpha$-viscosity model of \citet{1973A&A....24..337S} for which $\tnu=\alpha c_{\mathrm{s}}h$ with typical values of $\alpha$ of order $10^{-2}\text{--}10^{-1}$; here $c_\mathrm{s}$ is the sound speed and $h$ the scale height of the disk. With this parametrisation, the radial drift velocity from Eq.~(\ref{eq:ang-mom-cons-disk}) is
\begin{equation}
v_{\mathrm{r}}=-\frac{3}{2}\frac{\tnu}{r}=-\frac{3}{2}\frac{\alpha c_{\mathrm{s}}h}{r}=-\frac{3}{2}\alpha\left(\frac{h}{r}\right)^{2}v,
\label{eq:radial-velocity}
\end{equation}
where we have used $h=c_{\mathrm{s}}/\Omega$ and $v=r \Omega$. From mass conservation, we obtain the radial mass flux and thereby the mass-accretion rate onto the central star,
\begin{equation}
\dot{M}_{\mathrm{acc}} = 2\pi rh\rho |v_{\mathrm{r}}| = 3\pi\alpha c_{\mathrm{s}}^{2}\frac{\Sigma}{\Omega} \approx 3 \alpha \left(\frac{h}{r}\right)^2 \Omega M_\mathrm{disk}.
\label{eq:disk-mdot}
\end{equation}
In this equation, $\rho=\Sigma/h$ is the mass density and, in the last step, we have used that the total disk mass is approximately $M_\mathrm{disk} = \pi r^2 \Sigma$. For disks relevant in this work, this gives
\begin{equation}
\dot{M}_{\mathrm{acc}}=2.5\,\mathrm{M}_{\odot}\,\yr^{-1}\,\left(\frac{\alpha}{10^{-2}}\right)\left(\frac{h/r}{0.1}\right)^{2}\left(\frac{\Omega}{\mathrm{h}^{-1}}\right)\left(\frac{M_{\mathrm{disk}}}{\mathrm{M}_{\odot}}\right).
\label{eq:disk-mdot-typ-values}
\end{equation}

In the presence of a strong stellar magnetic field, a magnetosphere might form if the magnetic pressure overcomes the ram pressure of free-falling matter, that is if $B^2/8\pi \geq 1/2\rho v_\mathrm{ff}^2$. The mass-accretion rate of matter freely falling onto a central star of mass $M$ with velocity $v_\mathrm{ff}=\sqrt{2GM/r}=\sqrt{2}v_\mathrm{K}$ ($v_\mathrm{K}$ being the Keplerian velocity) and density $\rho$ is $\dot{M}=4\pi r^2 \rho v_\mathrm{ff}$. Then the magnetospheric radius $R_{\mathrm{m}}$ is given by
\begin{equation}
R_{\mathrm{m}}=\mu_\mathrm{B}^{4/7}\dot{M}^{-2/7}\left(2GM\right)^{-1/7},
\label{eq:Rm}
\end{equation}
where $\mu_\mathrm{B}$ is the magnetic dipole moment of the assumed dipolar stellar magnetic field (Eq.~\ref{eq:dipole-b-field}, Sect.~\ref{sec:b-field}). If this magnetospheric radius is larger than the stellar radius $R_{*}$, matter is deflected by the magnetic field and the accretion disk is effectively truncated at about $r=R_\mathrm{m}$. This locks the stellar surface to the angular frequency of the disk at $r=R_\mathrm{m}$ and reduces the specific angular momentum of accreted material compared to the case when the accretion disk extends all the way down to the stellar surface. The specific angular momentum accreted by the central star then is
\begin{equation}
j_\mathrm{acc}=
\begin{cases}
\frac{2}{3}R_{*}^2 \Omega(r=R_\mathrm{m}) & \text{for } R_{*} < R_\mathrm{m}\vspace{0.1cm} \\
\frac{2}{3}R_{*}^2 \Omega(r=R_{*}) & \text{for } R_{*} \geq R_\mathrm{m}.
\end{cases}
\label{eq:j-macc}
\end{equation}
In the disk-locked state, the co-rotation radius is equal to the magnetospheric radius and the stellar surface spins with an angular frequency of
\begin{equation}
\frac{\Omega_{*}}{\Omega_{\mathrm{crit}}}=\left(\frac{R_{*}}{R_\mathrm{m}}\right)^{3/2}=\left(2GM\right)^{3/14}\mu_\mathrm{B}^{-6/7}\dot{M}^{3/7}R^{3/2}
\label{eq:spin-disk-lock}
\end{equation} 
in terms of the stellar critical Keplerian angular frequency $\Omega_\mathrm{crit}=\sqrt{GM/R_{*}^3}$.

\subsection{From \threed to \oned}\label{sec:3d-to-1d}

Because of the large amount of angular momentum in the merger ($2 \times 10^{53}\,\mathrm{g}\,\mathrm{cm}^2\,\mathrm{s}^{-1}$), a star-disk structure composed of a central spherical object surrounded by a torus of about $3\,\msun$ forms \citep{2019Natur.574..211S,Ohlmann+2019}. The torus is thermally supported such that its rotational velocity profile is sub-Keplerian. The central, $14\,\msun$ spherical merger remnant is in solid-body rotation with an angular velocity matching that of the inner part of the sub-Keplerian torus. The boundary layer rotates at 76\% of the critical Keplerian value. At the end of the \threed MHD simulation, the central star carries about 40\% and the torus about 60\% of the total angular momentum. Such configurations of a central star surrounded by a thick torus are a common feature of mergers of stars and are observed, \eg, in mergers of double white dwarfs \citep[\eg][]{1990ApJ...348..647B,1995ApJ...438..887R,1997ApJ...481..355S,2010Natur.463...61P} and neutron stars \citep[\eg][]{1992ApJ...401..226R,1994ApJ...431..742D,1999A&A...341..499R}.

As explained in \citet{2019Natur.574..211S}, the evolution of the merger remnant right after the coalescence is set by that of the torus and how fast it is accreted onto the central remnant. For this, the accretion timescale $\tau_\mathrm{acc}$ and the cooling timescale $\tau_\mathrm{cool}$ of the torus are of particular relevance. If cooling is fast ($\tau_\mathrm{cool}\ll\tau_\mathrm{acc}$), the torus will evolve into a thin disk. In contrast, if accretion is fast ($\tau_\mathrm{cool}\gg\tau_\mathrm{acc}$), the torus is rapidly accreted without significant cooling and is thereby transformed into a bloated, thermally-supported stellar envelope.

In order to accrete matter from the torus, angular momentum has to be transported outwards. Within an $\alpha$-disk model (Sect.~\ref{sec:mass-accretion}), the accretion timescale $\tau_\mathrm{acc}$ is equivalent to the turbulent-viscous timescale $\tau_\mathrm{visc}$ and is set by the mass-accretion rate, $\dot{M}_{\mathrm{acc}}$ (Eq.~\ref{eq:disk-mdot}), and the total mass in the torus, $M_{\mathrm{disk}}$, \citep[see also][]{2019Natur.574..211S}
\begin{equation}
\tau_{\mathrm{acc}} = \frac{M_{\mathrm{disk}}}{\dot{M}_{\mathrm{acc}}} = \frac{1}{3}\frac{r^{2}}{\alpha h^{2}\Omega} \approx 0.02\,\yr\,\left(\frac{10^{-2}}{\alpha}\right)\left(\frac{r/h}{2}\right)^{2}\left(\frac{\mathrm{h}^{-1}}{\Omega}\right).
\label{eq:tau-acc}
\end{equation}

The cooling timescale can be estimated by considering how much turbulent (viscous) heating is produced by releasing gravitational potential energy when matter moves inwards and how fast this energy is lost by (radiative) cooling. The merger remnant can at most radiate at its Eddington luminosity such that the cooling timescale may be approximated as \citep[see also][]{2019Natur.574..211S}
\begin{align}
\tau_{\mathrm{cool}} &= \frac{E_{\mathrm{grav}}}{f_{\mathrm{Edd}} L_{\mathrm{Edd}}} = \frac{}{f_{\mathrm{Edd}}} \frac{GM_{\mathrm{core}}M_{\mathrm{disk}}/R_{\mathrm{core}}}{4\pi G(M_{\mathrm{core}}+M_{\mathrm{disk}})c/\kappa} \nonumber \\
 & \approx  0.8\times10^{3}\,\yr\, \left(\frac{1+X}{1.7}\right)\left(\frac{M_{\mathrm{core}}M_{\mathrm{disk}}/\mathrm{M}_{\odot}}{M_{\mathrm{core}}+M_{\mathrm{disk}}}\right)\left(\frac{R_{\odot}}{R_{\mathrm{core}}}\right).
\label{eq:tau-cool}
\end{align}
Here, $M_\mathrm{core}$ and $R_\mathrm{core}$ are the mass and radius of the central star, respectively, $\kappa$ is the opacity for which we assume it is dominated by electron scattering (\ie $\kappa=0.2\,(1+X)\,\mathrm{cm}^{2}\,\mathrm{g}^{-1}$ with $X$ the hydrogen mass fraction), and $f_\mathrm{Edd}$ is set to 1 in the last step. For the merger remnant studied here ($M_{\mathrm{core}}=14\,\mathrm{M}_{\odot}$, $M_{\mathrm{disk}}=3\,\mathrm{M}_{\odot}$, $R_{\mathrm{core}}=3\text{--}4\,\mathrm{R}_{\odot}$), the cooling timescale is about $500\text{--}700\,\yr$, which is significantly longer than the accretion timescale of less than $1\,\yr$ from Eq.~(\ref{eq:tau-acc}). Most of the torus is therefore expected to be rapidly accreted and transformed into an extended envelope before cooling becomes important. A similar outcome was anticipated by \citet{2012ApJ...748...35S} for the merger remnant of two white dwarfs, which has been further substantiated by more detailed simulations of the viscous evolution of the white-dwarf merger remnant by \citet{2012MNRAS.427..190S}.

\begin{figure*}
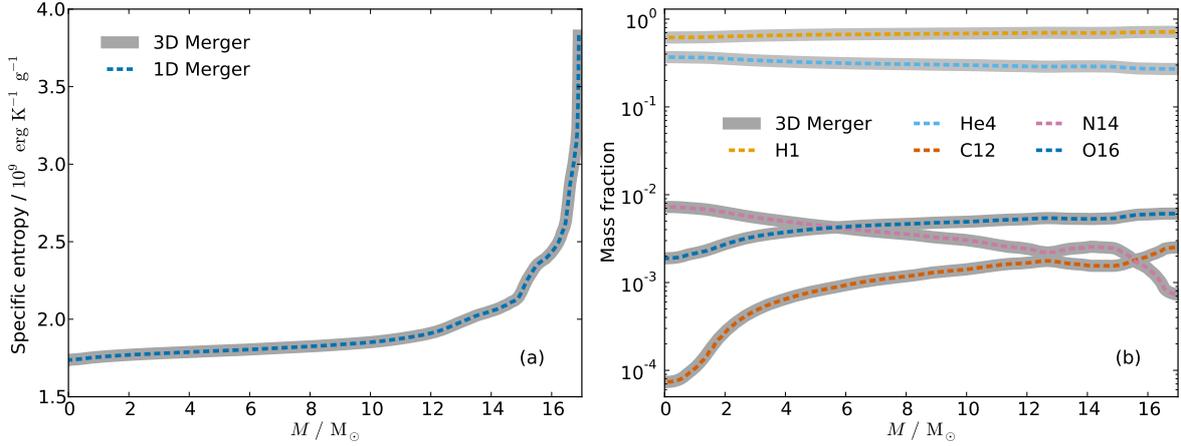

\begin{centering}
\includegraphics[width=0.9\textwidth]{{{init-condition-comparison-M-16.90}}}
\par\end{centering}
\vspace{-0.5cm}
\caption{Comparison of the initial interior profile of the \oned \mesa model and the \threed MHD merger remnant. In panel (a), we show specific entropy and in panel (b) we show the profiles of hydrogen (H1), helium (He4), carbon (C12), nitrogen (N14) and oxygen (O16) mass fractions. Figure reprinted with permission from \citet{2019Natur.574..211S}.}
\label{fig:imported-merger}
\end{figure*}

In \emph{Extended Data Fig.~1} of \citet{2019Natur.574..211S}, it is shown that the accretion and cooling timescales become comparable at a radius of about $54\,\rsun$, which corresponds to a mass coordinate of $16.9\,\msun$. It is therefore assumed that the inner $16.9\,\msun$ of the merger remnant forms a spherically symmetric star and that the remaining outer part of the torus cools and evolves into a (thin) disk. This then sets the initial condition for our \oned stellar evolution models. As described in \citet{2019Natur.574..211S}, we use relaxation routines in \mesa \citep[a detailed technical description of the applied relaxation routines can be found in Appendix~B of][]{2018ApJS..234...34P} to obtain a \oned stellar model with the same entropy and chemical structure as the \threed MHD merger remnant. After relaxation, the \oned structures closely match the \threed outcome (Fig.~\ref{fig:imported-merger}).

As described above, the inner spherical merger remnant is in solid-body rotation with an angular velocity that is set by the slightly sub-Keplerian velocity in the transition region between central merger remnant and outer torus \citep{2019Natur.574..211S,Ohlmann+2019}. During the viscous accretion of the torus, angular momentum is transported outwards and we assume that the inner star remains in solid-body rotation after most of the torus has been accreted. Such rotational profiles of rigidly rotating inner remnants that transition into sub-Keplerian disks are supported by more detailed simulations of the co-evolution of central white-dwarf merger remnants and a surrounding torus \citep{2012MNRAS.427..190S,2013ApJ...773..136J}. For the initial angular velocity of our \oned stellar models, we therefore assume an analogous rotational profile with a surface angular velocity set to 90\% of the critical Keplerian value.

\section{Post-merger evolution}\label{sec:merger-evolution}

\begin{figure}
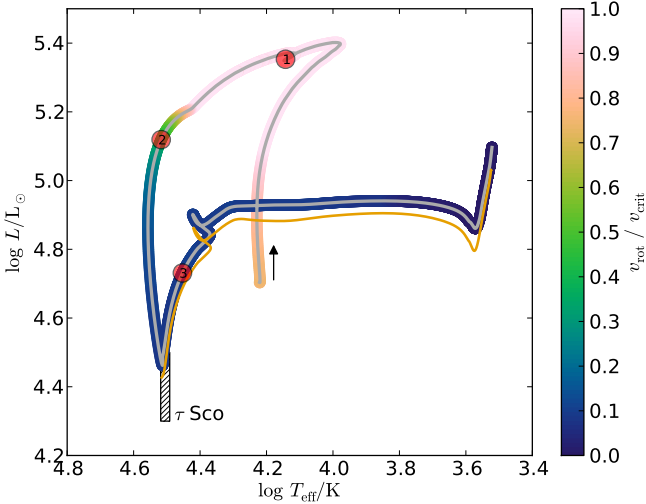

\begin{centering}
\includegraphics[width=0.49\textwidth]{{{hrd-M-16.90-mu-2e37-no-gradT-excess}}}
\par\end{centering}
\caption{Evolution of the \oned merger model with an assumed magnetic dipole moment of $\mu_\mathrm{B}=2\times10^{37}\,\mathrm{G}\,\mathrm{cm}^3$ in the Hertzsprung--Russell diagram. The surface rotation $v_\mathrm{rot}/v_\mathrm{crit}$ is colour coded and the points of interest 1, 2 and 3 in the evolution are marked for which some internal profiles are shown in Figs.~\ref{fig:profiles-poi-1}, \ref{fig:profiles-poi-2} and~\ref{fig:profiles-poi-3}, respectively. The arrow indicates the beginning and direction of evolution after accretion of most of the torus. The observed position of the magnetic massive star \tausco is indicated by the hatched box. For comparison, we also show the evolutionary track of an initially $16.9\,\msun$, genuine single star (orange line).}
\label{fig:hrd-mu-2e37}
\end{figure}

The evolution of our default merger model with a magnetic dipole moment of $\mu_\mathrm{B}=2\times 10^{37}\,\mathrm{G}\,\mathrm{cm}^3$ in the Hertzsprung--Russell (HR) diagram is shown in Fig.~\ref{fig:hrd-mu-2e37}. The equivalent diagrams for the merger model with a weaker and stronger magnetic field are presented in Appendix~\ref{sec:evol-different-b-fields} (Figs.~\ref{fig:hrd-mu-2e34} and~\ref{fig:hrd-mu-1e40}, respectively). The model starts as a ${\approx}\,30\,\rsun$ fast rotator with an effective temperature of $T_\mathrm{eff} \approx 16,000\,\mathrm{K}$ and a luminosity of $\log L/\lsun \approx 4.7$. Because of the merger, the model is out of thermal equilibrium and a rapid expansion phase sets in that is driven by the energy deposited in the star during the merging process. The star reaches a maximum luminosity of about $\log L/\lsun = 5.4$ and minimum effective temperature of about $T_\mathrm{eff} = 9700\,\mathrm{K}$ in about $2\,\yr$. The star never exceeds and also does not get close to its Eddington luminosity at the surface ($L/L_\mathrm{edd}<0.7$).

During the expansion phase to maximum luminosity, the surface reaches critical rotation. When the star rotates near break-up, we assume that it will shed as much mass as needed to bring it back to below critical rotation. During this phase, we do not consider magnetic braking and the shed mass is assumed to be lost from the star (${<}\,0.01\,\msun$ in this case which takes away about 7\% of the star's total angular momentum, see below). 

After the expansion phase, the merger contracts towards the MS while regaining thermal equilibrium. The contraction phase lasts for a few $10^3\,\yr$ and the surface spins down to about $50\,\kms$. We describe the spin evolution and the reasons for the spin-down in detail in Sect.~\ref{sec:spin-evolution}.

Once the star has reached thermal equilibrium, it continues its MS life and evolves similarly to a genuine single star of the same total mass (\cf the $16.9\,\msun$ single star in Fig.~\ref{fig:hrd-mu-2e37}). The main difference in the internal structure is that the merger model has a helium-enriched envelope. On average, the helium mass fraction in the envelope of the merger model is ${\approx}0.30$ while that of the comparison star is ${\approx}0.27$. This translates into mean molecular weights $\mu$ of ${\approx}0.63$ and ${\approx}0.61$, respectively, which is responsible for the slightly larger luminosity of the merger model on the MS because of the mass-luminosity relation \citep[$L\propto\mu^4 M^3$; \eg][]{1994sse..book.....K}. So in the case of the merger of relatively unevolved stars as studied here, the merger product adjusts its internal structure to its new mass and then continues its evolution closely to that of a genuine single star of the same mass.

In the beginning of its continued MS evolution (about $10^5\,\yr$ up to a few $10^6\,\yr$ after the merger), the merger product has an effective temperature ($T_\mathrm{eff}\approx 32500\,\mathrm{K}$), luminosity ($\log L/\lsun \approx 4.5$) and surface gravity ($\log g \approx 4.17$) that are in good agreement with the observed values of the magnetic star \tausco \citep[$T_\mathrm{eff}\approx 32000\,\mathrm{K}$, $\log L/\lsun \approx 4.3\text{--}4.5$ and $\log g \approx 4.0\text{--}4.3$;][]{2005A&A...441..711M,2006A&A...448..351S,2014A&A...566A...7N}. Therefore our merger model similarly stands out as a blue straggler in the Upper Scorpius association just as \tausco does. 

Our merger model is a slow rotator, albeit rotating faster than \tausco, which has a projected rotational velocity of ${\lesssim}\,15\,\kms$ \citep[\eg][]{2006A&A...448..351S}, and it is somewhat over-luminous. The luminosity match with \tausco could easily be improved by considering the merger of slightly less massive stars. The surface of our merger model is not enriched in nuclear burning products, in contrast to what is observed in \tausco. However, the average nitrogen mass fraction in the envelope of the merger product is $0.0025$, which corresponds to an enrichment of a factor of almost 3.5 compared to the initial nitrogen mass fraction of $0.00074$ used in our computations. So slightly more mixing of the envelope or some additional mass loss could significantly enrich the surface of our merger model (see Sect.~\ref{sec:discussion-surface-enrichment}). Mixing in particular may have been underestimated in our modelling because we did not consider any mixing during the accretion of the torus and assumed that the strong magnetic field does not contribute to chemical mixing. In conclusion, our merger model appears compatible with the observational properties of \tausco and therefore provides a promising way to explain this enigmatic, magnetic blue straggler.

The evolution of the less and more magnetic merger models are very similar to that of the default model (Appendix~\ref{sec:evol-different-b-fields}). The main difference is that it takes the less magnetic model more time to spin down to about the same surface rotational velocity as the default model and that the merger model with the stronger magnetic field spins even more slowly after thermal relaxation. The latter is discussed in more details in Sect.~\ref{sec:spin-evolution}.

\subsection{Spin evolution}\label{sec:spin-evolution}

\begin{figure}
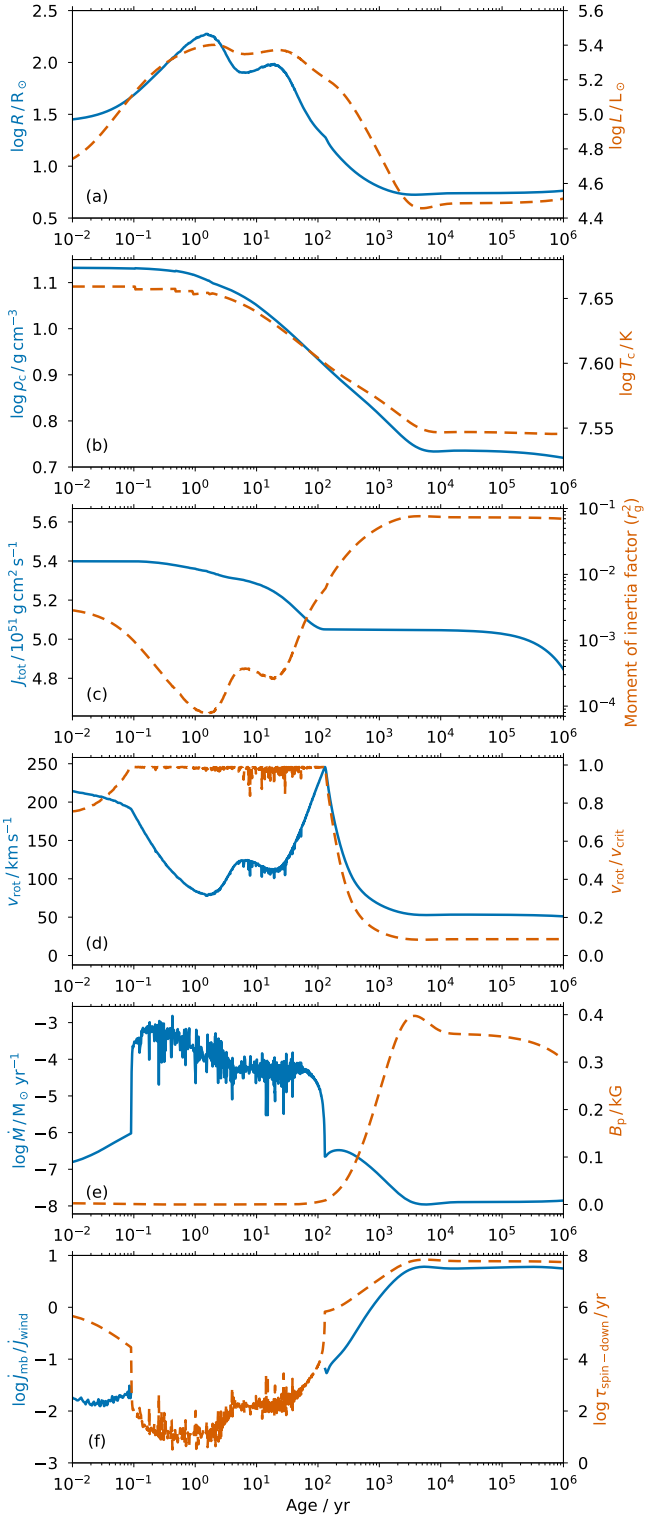

\begin{centering}
\includegraphics[width=0.49\textwidth]{{{time-evol-M-16.90-mu-2e37-no-gradT-excess}}}
\par\end{centering}
\caption{Evolution of (a) stellar radius $\log R/\rsun$ and luminosity $\log L/\lsun$, (b) core density $\log \rho_\mathrm{c}$ and core temperature $\log T_\mathrm{c}$, (c) total angular momentum $\log J_\mathrm{tot}$ and moment of inertia factor $r_\mathrm{g}^2$, (d) surface rotational velocity $v_\mathrm{rot}$ and fraction of critical surface rotation $v_\mathrm{rot}/v_\mathrm{crit}$, (e) mass-loss rate $\log \dot{M}$ and surface polar magnetic field strength $B_\mathrm{p}$, and (f) ratio of angular-momentum loss through magnetic braking and wind mass loss $\dot{J}_\mathrm{mb}/\dot{J}_\mathrm{wind}$ and spin-down timescale $\tau_\mathrm{spin-down}$ of the \oned merger model. The magnetic dipole moment of this model is $\mu_\mathrm{B}=2\times10^{37}\,\mathrm{G}\,\mathrm{cm}^3$.}
\label{fig:time-evol-mu-2e37}
\end{figure}

The default merger model rapidly spins down from near critical rotation right after the merger to about $50\,\kms$ after thermal relaxation. This spin-down is not because of angular-momentum loss from the star but rather because of a low initial angular-momentum content of the star. To illustrate this, we show the evolution of various quantities of the merger product in Fig.~\ref{fig:time-evol-mu-2e37} (equivalent plots for the merger models with weaker and stronger magnetic field are in Figs.~\ref{fig:time-evol-mu-2e34} and~\ref{fig:time-evol-mu-1e40}, respectively). The top panel (Fig.~\ref{fig:time-evol-mu-2e37}a) shows the radius and luminosity evolution and allows us to connect different phases of the thermal relaxation with, \eg, the position of the star in the HR diagram (Fig.~\ref{fig:hrd-mu-2e37}).

The merger product can only lose angular momentum through mass loss and magnetic braking. The star reaches critical rotation during the expansion of the envelope as shown in Figs.~\ref{fig:hrd-mu-2e37} and~\ref{fig:time-evol-mu-2e37}d. During this phase, less than $0.01\,\msun$ (\ie $<0.1\%$ of the total mass) are shed from the star with an average rate of order $10^{-4}\,\msun\,\yr^{-1}$ (Fig.~\ref{fig:time-evol-mu-2e37}e). As is evident from Fig.~\ref{fig:time-evol-mu-2e37}c, this mass loss removes angular-momentum of about $0.4\times10^{50}\,\mathrm{g}\,\mathrm{cm}^2\,\mathrm{s}^{-1}$, corresponding to about 7\% of the total initial angular momentum of the star. Such a relatively small loss cannot explain the spin-down of the star from nearly critical rotation.

Magnetic braking is not considered while the star rotates critically and sheds mass because of numerical reasons. In any case, it would be inefficient in this phase because of the large radius and hence small surface magnetic field due to the dipole scaling (Fig.~\ref{fig:time-evol-mu-2e37}e). Compared to the angular-momentum loss through stellar winds alone, the contribution of magnetic braking is negligible up to about $10^3\,\yr$. Only then does magnetic braking exceed the angular-momentum loss from only stellar winds (Fig.~\ref{fig:time-evol-mu-2e37}f). Throughout the evolution of the merger model, the spin-down timescale (Eq.~\ref{eq:spin-down-timescale}) is long and is about $50\text{--}60\,\myr$ once the star finished its thermal relaxation and evolves along the MS (Fig.~\ref{fig:time-evol-mu-2e37}f). So the magnetic field in this model is negligible in terms of angular-momentum loss but it is important regarding the angular-momentum transport through the interior of the star and enables solid-body rotation as we will show in Sect.~\ref{sec:interior-evolution} below.

The observed spin-down of the star is thus not caused by angular-momentum loss but rather by an internal restructuring. The surface rotational velocity drops from about $250\,\kms$ to $50\,\kms$ from $10^2$ to $10^4\,\yr$ after the merger (Fig.~\ref{fig:time-evol-mu-2e37}d). At the same time, the radius decreases by about a factor of 4 and $r_\mathrm{g}^2$ increases by about a factor of 20. The surface velocity of a star in solid-body rotation and constant mass and angular momentum evolves according to $v=R\Omega=J/r_\mathrm{g}^2 M R\propto (r_\mathrm{g}^2 R)^{-1}$. So the surface velocity is expected to decrease by a factor of 5, which fully explains the observed spin-down of the merger product. During this evolution, the surface angular velocity $\Omega$ and moment of inertia stay almost constant, showing once more that the spin-down is not because of angular-momentum loss.

\begin{figure}
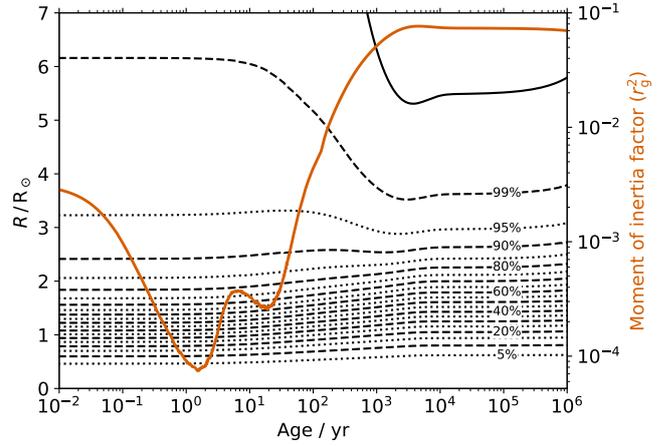

\begin{centering}
\includegraphics[width=0.49\textwidth]{{{internal-restruct-M-16.90-mu-2e37-no-gradT-excess}}}
\par\end{centering}
\caption{Internal mass readjustment of the \oned merger model. The black dashed and dotted lines show the radial location of various mass coordinates (in steps of 5\% as indicated by the labels) and the red solid line the moment of inertia factor from Fig.~\ref{fig:time-evol-mu-2e37}c. The black solid line indicates the stellar surface.}
\label{fig:internal-restruct}
\end{figure}

The change of $r_\mathrm{g}^2$ from $10^2$ to $10^4\,\yr$ after the merger is caused by two processes, by the contraction of the envelope and the expansion of the core. After the merger, the star's core density and core temperature are larger and hotter than in thermal equilibrium. This drives enhanced nuclear burning and thereby leads to an expansion and cooling of the core region towards the equilibrium state (Fig.~\ref{fig:time-evol-mu-2e37}b). The expansion of the core and contraction of the envelope are illustrated in Fig.~\ref{fig:internal-restruct} by the temporal evolution of the radius of various mass shells. Only the outermost ${\approx}\,5\%$ in mass take part in the overall expansion and later contraction of the star while the innermost ${\approx}\,90\%$ in mass expand and evolve into a less centrally-concentrated density configuration. The simultaneous changes of core and envelope leave the moment of inertia nearly constant such that the change in $r_\mathrm{g}^2=I/MR^2$ is mainly driven by the change in radius.

As shown above, the spin-down of the star is caused by internal restructuring of the star. Because of the small initial moment of inertia factor of the star after the merger ($r_\mathrm{g}^2\approx 3\times10^{-3}$, Fig.~\ref{fig:time-evol-mu-2e37}c), the total angular momentum of a solid-body rotator near break-up is small,
\begin{equation}
J = r_\mathrm{g}^2 M R^2 \Omega_\mathrm{crit} \propto r_\mathrm{g}^2 M^{3/2} R^{1/2}.
\label{eq:init-ang-mom}
\end{equation}
Even if the radius of the merger product was larger by a factor of 2, its angular momentum would only increase by about 40\%. In our case, the total initial angular momentum of the model is $5.4\times10^{51}\,\mathrm{g}\,\mathrm{cm}^2\,\mathrm{s}^{-1}$, while the merger remnant up to $16.9\,\msun$ before the viscous evolution contained ${\approx}\,2\times10^{53}\,\mathrm{g}\,\mathrm{cm}^2\,\mathrm{s}^{-1}$. In other words, our \oned model only contains about 3\% of the available angular momentum despite rotating near break-up at the surface. By assuming that the merger product rotates rigidly after the accretion of the torus, we have implied an efficient angular-momentum transport not only in the torus but also in the central merger remnant (see Sect.~\ref{sec:3d-to-1d}).

The details of the viscous evolution of the star-torus structure right after the coalescence are uncertain at the moment, especially how much angular momentum is transported out of the star into the remaining torus/disk. We therefore also consider the case that ten times more angular momentum ($5\times10^{52}\,\mathrm{g}\,\mathrm{cm}^2\,\mathrm{s}^{-1}$) remains in the merger product than considered before. In this case, the star can initially not be in solid-body rotation but it has to rotate differentially. We impose a rotation profile with constant specific angular momentum. In fact, the initial distribution of angular momentum in the star is not essential, because the magnetic fields quickly redistribute the angular momentum and try to approach solid-body rotation.

\begin{figure}
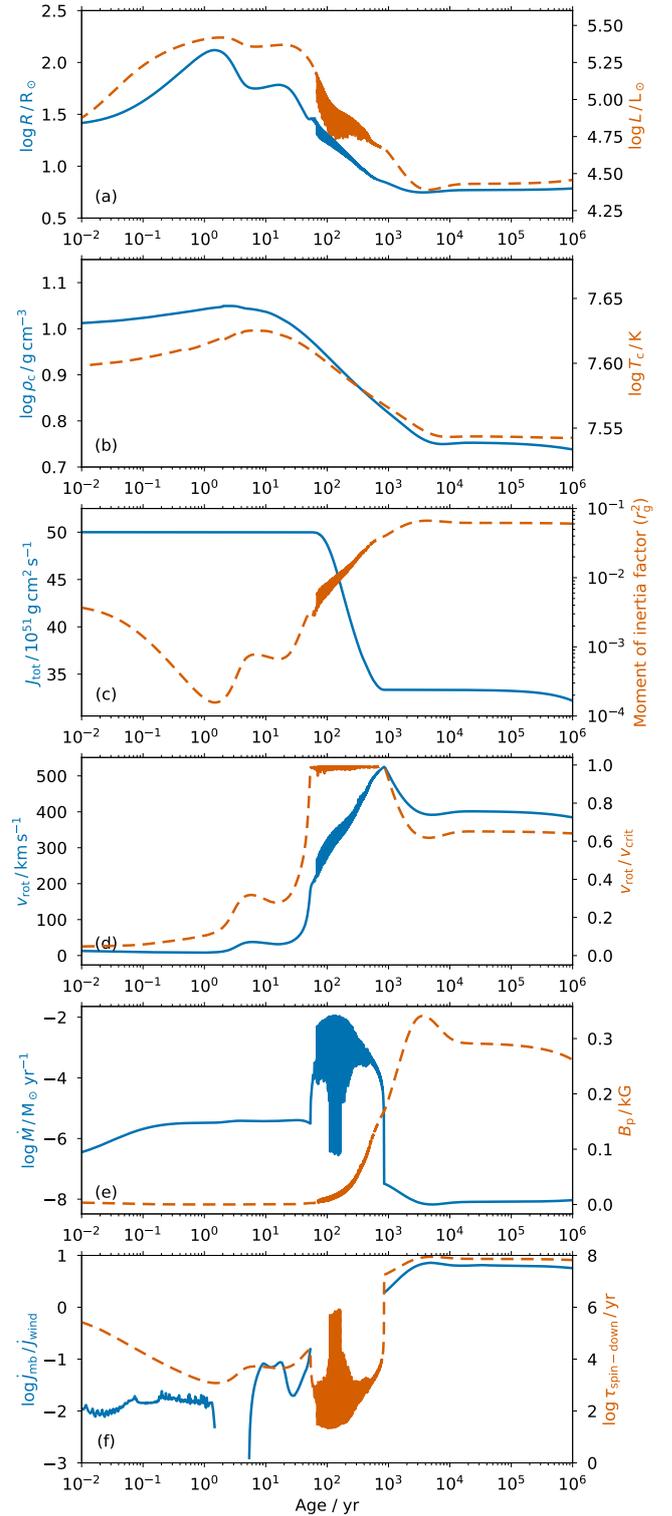

\begin{centering}
\includegraphics[width=0.49\textwidth]{{{time-evol-M-16.90-mu-2e37-J-5e52-no-gradT-excess}}}
\par\end{centering}
\caption{As Fig.~\ref{fig:time-evol-mu-2e37} but for a merger model with a ten times larger initial angular momentum, $5\times10^{52}\,\mathrm{g}\,\mathrm{cm}^2\,\mathrm{s}^{-1}$. The assumed magnetic dipole moment is $\mu_\mathrm{B}=2\times10^{37}\,\mathrm{G}\,\mathrm{cm}^3$ as before.}
\label{fig:time-evol-mu-2e37-J-5e52}
\end{figure}

Details of the evolution of this model are shown in Fig.~\ref{fig:time-evol-mu-2e37-J-5e52}. Because it is challenging to model this star numerically, the mass-loss rate to keep the star below critical rotation (Fig.~\ref{fig:time-evol-mu-2e37-J-5e52}e) and the luminosity oscillate (Fig.~\ref{fig:time-evol-mu-2e37-J-5e52}e) more than in the model with less initial angular momentum (Fig.~\ref{fig:time-evol-mu-2e37}). The star now reaches critical rotation later in its evolution after the angular momentum has been transported from the interior to the surface (because of the different initial rotational velocity profile, most of the angular momentum is now deep in the star and no longer near the surface). At critical rotation, the star now loses $0.5\,\msun$ and about 35\% of its angular momentum, and the angular-momentum loss timescale in this phase is set by that of angular-momentum transport from the interior to the surface. If this transport was more efficient, more mass and more angular momentum would have been lost. This can be seen by comparing our above default model to a model with a stronger magnetic field (Figs.~\ref{fig:hrd-mu-1e40} and~\ref{fig:time-evol-mu-1e40}). The mass- and hence angular-momentum-loss rates are larger because the angular-momentum transport to the surface is faster (the mass-loss rate is about one order of magnitude larger). The model with the stronger magnetic field loses roughly half of its initial angular momentum compared to only 7\% in our default model, an increase by a factor of 7. Increasing the magnetic field further, could even allow the star to lose all its angular momentum in this phase.

With the larger initial angular momentum, the merger model rotates with ${\approx}\,400\,\kms$ at the surface after thermal relaxation, that is with roughly 65\% of break-up (Fig.~\ref{fig:time-evol-mu-2e37-J-5e52}d). So this merger model would be a rapid rotator. Because of the stronger mass loss, the model exposes a nitrogen enriched surface (mass fraction of about 0.0011, \ie an increase of about 50\% compared to the initial nitrogen abundance) that continuously becomes more enriched through rotational mixing. For the chosen magnetic field strength, magnetic braking is unimportant and the spin-down time exceeds the MS lifetime of the star (Fig.~\ref{fig:time-evol-mu-2e37-J-5e52}f). In case of a stronger magnet field, magnetic braking would become relevant.

\subsection{Interior evolution and post-merger rejuvenation}\label{sec:interior-evolution}

\begin{figure}
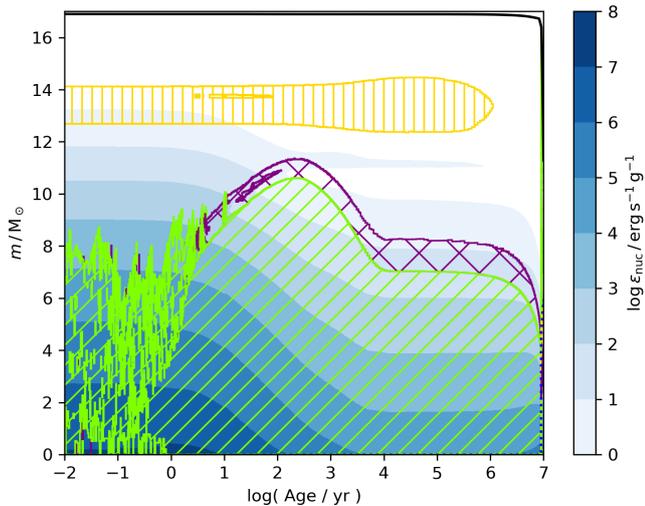

\begin{centering}
\includegraphics[width=0.49\textwidth]{{{kippenhahn-M-16.90-mu-2e37-no-gradT-excess}}}
\par\end{centering}
\caption{Kippenhahn diagram of the internal evolution of the \oned merger model up to core-hydrogen exhaustion. The green, purple and yellow hatchings indicate the occurrence of convection, convective-core overshooting and thermohaline mixing, respectively. The blue color bar shows nuclear energy generation, $\epsilon_\mathrm{nuc}$.}
\label{fig:kippenhahn}
\end{figure}

We now return to our default model and highlight a few features of its internal evolution. As shown in Fig.~\ref{fig:time-evol-mu-2e37}b, the core of the star after the merger is denser and hotter than in full equilibrium. In the merger process, energy has been deposited in the interior of the star that causes a thermal imbalance and drives the evolution. Also, the carbon (C) and nitrogen (N) abundances in the core are no longer in CN equilibrium\footnote{Note that this is similar to what can happen during the pre-main-sequence evolution, where CN equilibrium is achieved before the onset of core-hyrodgen burning and a transient convective core can be driven.} because of mixing in the merger (Fig.~\ref{fig:imported-merger}b). Together, the larger density and higher temperature as well as the non-CN equilibrium abundances, result in enhanced nuclear burning and a transient, unusually large convective core forms as can be seen in Fig.~\ref{fig:kippenhahn}. The convective core reaches a maximum mass extent of about $11\,\msun$ at ${\approx}300\,\yr$ ($\log t/\yr\approx 2.5$) while the mass of the convective core in equilibrium is only about $8\,\msun$. This short phase with an enlarged convective core mixes fresh hydrogen into the core of the merger product and thereby further contributes to its rejuvenation. It also leaves a chemical gradient in the envelope (see below).

Rejuvenation is at the heart of the blue straggler phenomenon and always occurs when a star accretes mass, for example via mass transfer in Roche-lobe filling binaries or via mergers. It is caused by two effects: (i) the apparent rejuvenation due to shorter lifetimes associated with more massive stars after mass gain and (ii) the mixing of fresh fuel into the cores of stars that prolongs their lifetimes \citep[see \eg][]{2016MNRAS.457.2355S}. In Table~\ref{tab:rejuvenation}, we report the core hydrogen and helium mass fractions of the $9\,\msun$ primary and $8\,\msun$ secondary progenitors and the merger products right after coalescence and ${\approx}\,10^4\,\yr$ after the beginning of the thermal relaxation. With respect to the primary star, the core hydrogen mass fraction increases by 9\% during the merger process and by in total 12\% after the additional mixing from the large transient convective core during the thermal relaxation of the merger product. So a substantial fraction of the rejuvenation takes place later during the relaxation phase and not exclusively in the merger process itself. In our case, the overall mixing of fresh fuel into the core is modest because the progenitor stars are rather unevolved (they are roughly 30\% through their MS evolution). For more evolved progenitors, the mixing of fresh fuel into the core is expected to be larger \citep[\eg][]{2016MNRAS.457.2355S}, but the inhibition of such mixing through strong chemical gradients is uncertain and needs further detailed modelling.

\begin{table}
\caption{\label{tab:rejuvenation}Average core hydrogen, $\left<X_\mathrm{H}\right>$, and core helium, $\left<X_\mathrm{He}\right>$, mass fractions of the $9\,\msun$ primary and $8\,\msun$ secondary progenitors, the merger product directly after the coalescence and the merger product at ${\approx}\,10^4\,\yr$ after coalescence. The average is computed over the inner $2\,\msun$. The relative increase in core hydrogen mass fraction is with respect to the primary, secondary and merger product after coalescence.}
\centering
\begin{tabular}{lccccc}
\toprule 
Star & $\left<X_\mathrm{H}\right>$ & $\left<X_\mathrm{He}\right>$ & \multicolumn{3}{c}{Rel.\ $\left<X_\mathrm{H}\right>$ increase}\\
 & & & Prim. & Sec. & Merger \\
\midrule
\midrule 
Primary ($9\,\msun$) & 0.5908 & 0.3952 & -- & -- &-- \\
Secondary ($8\,\msun$) & 0.6169 & 0.3690 & -- & -- & -- \\
$0\,\yr$ after merger & 0.6438 & 0.3420 & 9\% & 4\% & -- \\
$10^4\,\yr$ after merger & 0.6600 & 0.3257 & 12\% & 7\% & 2.5\% \\
\bottomrule
\end{tabular}
\end{table}

The Kippenhahn diagram in Fig.~\ref{fig:kippenhahn} further reveals a region of thermohaline mixing. This is where a considerable fraction of the core of the primary star ended up when it was tidally disrupted \citep{2019Natur.574..211S}. This material from the former primary star has a larger mean molecular weight than the deeper layers and thus causes the mixing. The core material of the merger product is primarily made of the core of the secondary star.

\begin{figure}
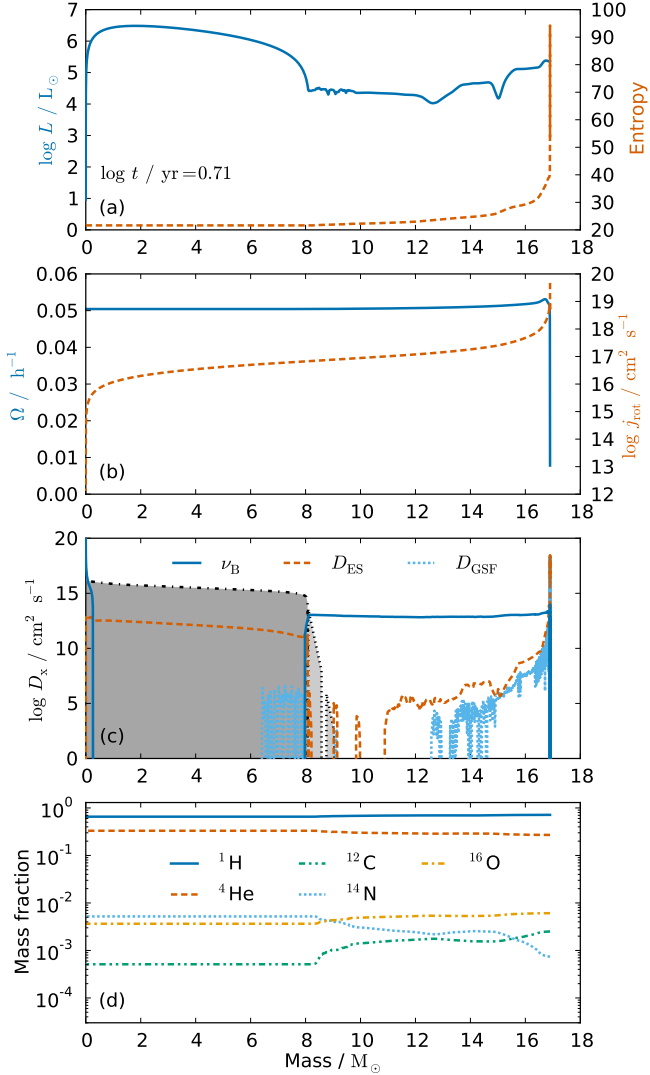

\begin{centering}
\includegraphics[width=0.49\textwidth]{{{profiles-M-16.90-mu-2e37-no-gradT-excess-pn-47}}}
\par\end{centering}
\vspace{-1cm}
\caption{Interior profiles of the \oned merger model with an assumed magnetic dipole moment of $\mu_\mathrm{B}=2\times10^{37}\,\mathrm{G}\,\mathrm{cm}^3$ after ${\approx}\,5\,\mathrm{yr}$ of the merger (point~1 in the HR diagram, Fig.~\ref{fig:hrd-mu-2e37}). The profiles show (a) the internal luminosity $\log L/\lsun$ and entropy, (b) the angular velocity $\Omega$ and specific angular momentum $\log j_\mathrm{rot}$, (c) the effective magnetic viscosity $\tnu_\mathrm{B}$ and Eddington--Sweet $D_\mathrm{ES}$ and Goldreich--Schubert--Fricke $D_\mathrm{GSF}$ diffusion coefficients and (d) the hydrogen, helium, carbon, nitrogen and oxygen mass fractions. In panel (c), the grey shaded regions indicate the convective core and assumed overshooting regions with their respective diffusion coefficients.}
\label{fig:profiles-poi-1}
\end{figure}

\begin{figure}
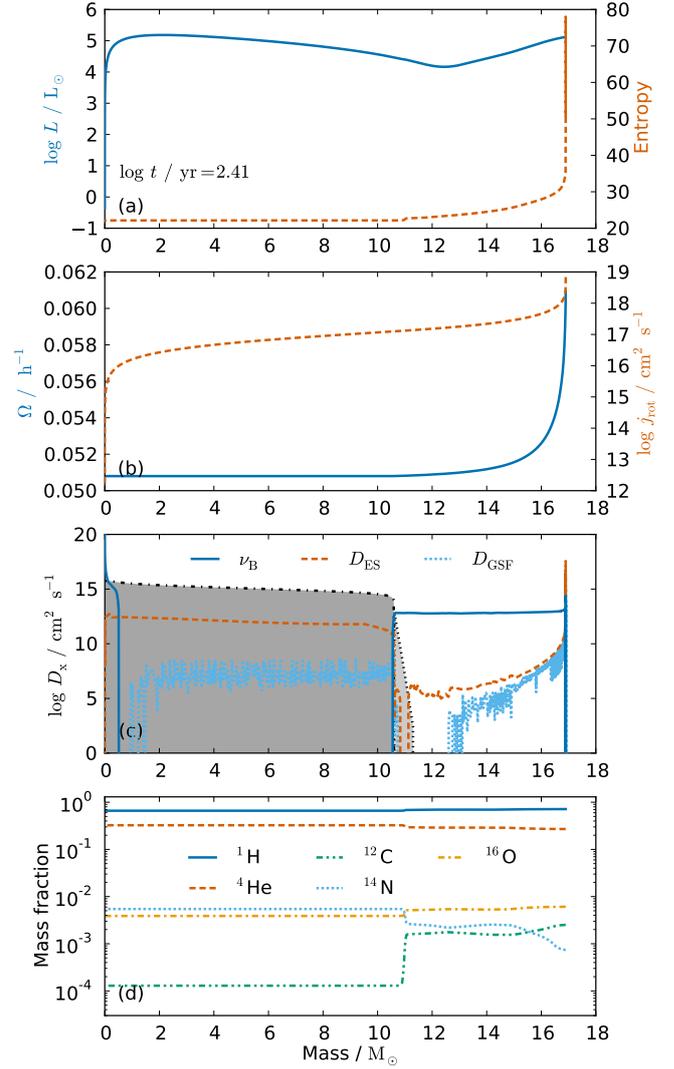

\begin{centering}
\includegraphics[width=0.49\textwidth]{{{profiles-M-16.90-mu-2e37-no-gradT-excess-pn-95}}}
\par\end{centering}
\vspace{-1cm}
\caption{As Fig.~\ref{fig:profiles-poi-1} but ${\approx}\,260\,\mathrm{yr}$ after the merger (point~2 in the HR diagram, Fig.~\ref{fig:hrd-mu-2e37}).}
\label{fig:profiles-poi-2}
\end{figure}

In Figs.~\ref{fig:profiles-poi-1}, \ref{fig:profiles-poi-2} and~\ref{fig:profiles-poi-3}, we show internal profiles of the merger product around the time of maximum luminosity, maximum convective core size and the later phase of the MS evolution (points 1, 2 and 3 in Fig.~\ref{fig:hrd-mu-2e37} corresponding to $5\,\yr$, $260\,\yr$ and $6.9\,\myr$ after coalescence, respectively). In particular, we focus on the luminosity, rotation, angular-momentum transport and chemical profiles. The thermal imbalance of the model is best seen in the luminosity profiles -- once in thermal equilibrium, the profiles are flat such that the luminosity produced in the core equals that on the surface. Early in the relaxation process (Figs.~\ref{fig:profiles-poi-1}a and~\ref{fig:profiles-poi-2}a), this is not the case and the core luminosity reaches values of up to $\log L/\lsun=6.4$, exceeding the maximum surface luminosity during relaxation by one order of magnitude. This large core luminosity drives the expansion of the core region and the large convective core size.

The angular rotational velocity profile $\Omega$ is always close to solid-body rotation, but less so in the beginning in the outermost layers that take part in the initial expansion (surface rotates slower than interior; Fig.~\ref{fig:profiles-poi-1}b) and later contraction (faster surface rotation than in the interior; Fig.~\ref{fig:profiles-poi-2}b). Once on the MS, the timescale on which the radius of the star changes (nuclear timescale) is much slower than the timescale of angular-momentum transport through the large-scale magnetic field (\alfven timescale) and the interior rotational profile closely matches that of a solid-body rotator (Fig.~\ref{fig:profiles-poi-3}b).

In panels (c) of Figs.~\ref{fig:profiles-poi-1}, \ref{fig:profiles-poi-2} and~\ref{fig:profiles-poi-3}, we show the diffusion coefficients of several mixing processes. The convective core and overshooting regions are indicated by the gray shaded areas and we show the diffusion coefficient of angular-momentum transport through the large-scale magnetic field ($\tnu_\mathrm{B}$, Eq.~\ref{eq:nu-eff}) in comparison to those of Eddington--Sweet circulations \citep[$D_\mathrm{ES}$;][]{1925Obs....48...73E,1950MNRAS.110..548S,1974IAUS...66...20K,1978ApJ...220..279E,2000ApJ...528..368H} and the Goldreich--Schubert--Fricke instability \citep[$D_\mathrm{GSF}$;][]{1967ApJ...150..571G,1968ZA.....68..317F,2000ApJ...528..368H}. The effective magnetic viscosity in the radiative envelope reaches values of $10^{13}\text{--}10^{14}\,\mathrm{cm}^2\,\mathrm{s}^{-1}$. From the diffusion coefficients it is evident that magnetic mixing enforces solid-body rotation basically in the whole radiative envelope while convection enforces solid-body rotation in the core. The magnetic fields extend slightly into the convective-core overshooting region and thereby provide some coupling of the core and envelope. Moreover, because of the receding convective core during the MS evolution, the magnetic field couples formerly convective regions with the radiative envelope and thereby ensures solid-body rotation basically throughout the whole star. Right in the centre of our model, one can observe magnetic mixing because of the singular behaviour of the dipole magnetic field. The mixing does not affect the evolution and is an artefact of our modelling. At the surface, Eddington--Sweet circulation dominates the mixing in our models. Shear mixing is unimportant because of solid-body rotation and is therefore not shown.

\begin{figure}
\begin{centering}
\includegraphics[width=0.49\textwidth]{{{profiles-M-16.90-mu-2e37-no-gradT-excess-pn-130}}}
\par\end{centering}
\vspace{-1cm}
\caption{As Fig.~\ref{fig:profiles-poi-1} but ${\approx}\,6.9\,\mathrm{Myr}$ after the merger (point~3 in the HR diagram, Fig.~\ref{fig:hrd-mu-2e37}).}
\label{fig:profiles-poi-3}
\end{figure}

The surface abundances of the merger model are close to the initial values of the progenitors (Figs.~\ref{fig:profiles-poi-1}d, \ref{fig:profiles-poi-2}d and~\ref{fig:profiles-poi-3}d). This merger product would therefore not be able to explain some of the nitrogen enriched, slow rotators in the Hunter diagram \citep{2008ApJ...676L..29H,2011A&A...530A.116B,2012ARA&A..50..107L}. However, core material of the former primary star is mixed in the envelope at mass coordinates of $13\text{--}15\,\msun$, \ie close to the surface (Figs.~\ref{fig:profiles-poi-1}d). These layers cause the thermohaline mixing seen in Fig.~\ref{fig:kippenhahn} and further mass loss and/or additional chemical mixing during the thermal relaxation could easily expose them, as is for example the case in the merger model with the larger initial angular-momentum content discussed above. At the maximum extent of the convective core, the innermost $11\,\msun$ are fully mixed, thus supplying it with fresh fuel (Fig.~\ref{fig:profiles-poi-2}d). A left-over chemical gradient in the envelope develops at the maximum extent of the convective core (Fig.~\ref{fig:profiles-poi-3}d) and is present until the first dredge-up (not shown here). The chemical gradient is not very strong because the initial progenitors were relatively unevolved. If they were more evolved, they would have a larger helium content and the gradient would most likely be larger. It may even influence shell burning episodes and dredge-up when ascending the red supergiant branch.

Our merger model evolves into a red supergiant where the first dredge-up will remove the chemical gradient left by the transient convective core. In other merger cases without a red-supergiant phase \citep[\eg in Case~B mergers,][]{2014ApJ...796..121J}, the chemical gradient will not be eroded and may be relevant at the supernova stage.

\subsection{Fate of circumstellar disk material}\label{sec:fate-disk}

Not all material of the torus is accreted onto the central merger remnant. In our default model, there is about $0.1\,\msun$ left that we envision to form a thin disk because it can cool faster than it is expected to be accreted. In the following, we explore how much of that mass might be lost in a wind (Sect.~\ref{sec:disk-wind}) and how the late accretion of that material could influence the evolution of the merger product (Sect.~\ref{sec:reaccretion-disk}).

\subsubsection{Disk wind}\label{sec:disk-wind}

\begin{figure}
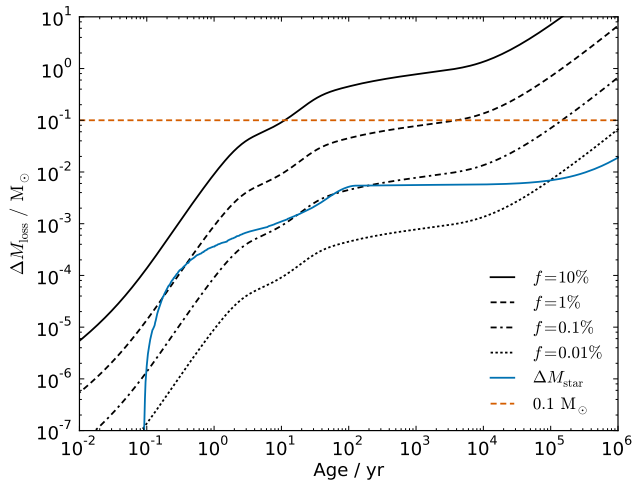

\begin{centering}
\includegraphics[width=0.49\textwidth]{{{mass-loss-M-16.90-mu-2e37-no-gradT-excess}}}
\par\end{centering}
\caption{Integrated mass loss of the \oned merger model, $\Delta M_\mathrm{star}$, and of a disk wind that is driven by a fraction $f$ of the total stellar luminosity. For reference, a total mass of $0.1\,\msun$ is highlighted.}
\label{fig:mass-loss}
\end{figure}

It is uncertain with which rate mass can be lost from a thin disk around the merger product. We here make the assumption that a fraction $f$ of the stellar luminosity is used to lift disk material out of the gravitational potential of the merger to infinity, thereby evaporating the disk. Hence,
\begin{equation}
f L_{*} \Delta t  =  \frac{GM_{*}\Delta M}{2R_{*}},
\label{eq:disk-wind-energy}
\end{equation}
such that the mass-loss rate of the disk is
\begin{align}
\dot{M}_\mathrm{disk} & = f \frac{2R_{*} L_{*}}{GM_{*}} \nonumber \\
 & \approx 3.15 \times 10^{-7} \,\msun\,\yr^{-1} \, f \left(\frac{R_{*}}{5\,\rsun}\right)\left(\frac{L_{*}}{\lsun}\right)\left(\frac{\msun}{M_{\mathrm{*}}}\right).
\end{align}
The resulting, cumulative mass-loss from a thin disk for our default model for various fractions of $f$ is shown in Fig.~\ref{fig:mass-loss}. 

It is evident that, if only a small fraction of the stellar luminosity ($f<1\%$) is used to drive disk mass-loss, all disk material can be lost and might never be accreted onto the merger product. Also the mass lost from the merger while it evolves at break-up velocity could energetically be easily lost as is shown by the comparison of the integrated disk wind and the total stellar mass loss of the model (Fig.~\ref{fig:mass-loss}). This implies that the merger product may not show a remnant disk after the thermal relaxation phase.

\subsubsection{Re-accretion onto merger remnant}\label{sec:reaccretion-disk}

If the remaining disk material is not lost from the star, it could instead be accreted. The accretion will mainly increase the angular momentum of the merger and thus its rotational velocity. If a star of mass $M$ with radius $R=R_\mathrm{acc}$ accretes from a Keplerian disk, the accreted specific angular momentum is $j_{\mathrm{acc}}=R_\mathrm{acc}^{2} \Omega_\mathrm{K}=\sqrt{G M R_\mathrm{acc}}$. Once on the MS, the star rotates as a solid body and its angular momentum is $J=I\Omega$. After accreting a mass of $\Delta M$, \ie angular momentum $J_{\mathrm{acc}}=\sqrt{G M R_{\mathrm{acc}}}\Delta M$, the new angular velocity is
\begin{equation}
\Omega=\frac{J_{0}+J_{\mathrm{acc}}}{I}=\Omega_{0}+\frac{J_{\mathrm{acc}}}{I}=\Omega_{0}+\Delta\Omega\,,
\label{eq:omega-after-accretion}
\end{equation}
where $J_0$ is the angular momentum of the star before mass accretion. The change of the angular velocity is
\begin{equation}
\Delta\Omega = \frac{\sqrt{G M R_{\mathrm{acc}}}}{I} \Delta M\,,
\label{eq:domega}
\end{equation}
meaning that it is linearly proportional to the amount of accreted mass (the change of the rotational velocity is $\Delta v = R\Delta \Omega$) and that more angular momentum is accreted for larger radii $R_{\mathrm{acc}}$.

We check and confirm the above analytic expectations by accreting a total of $\Delta M=0.1\,\msun$ at two different phases onto the merger models. We further take into account that the surface magnetic field may form a magnetosphere (\cf Eq.~\ref{eq:Rm}) such that mass is accreted with a lower specific angular momentum (\cf Eqs.~\ref{eq:Rm} and~\ref{eq:j-macc} in Sect.~\ref{sec:mass-accretion}). The mass is accreted onto our default model and the model with the larger magnetic field at luminosities of $\log L/\lsun = 4.7$ and $5.1$ with an accretion rate given by Eq.~\ref{eq:disk-mdot-typ-values} for $\alpha=0.01$ and $h/r=0.1$.

\begin{table}
\caption{\label{tab:reaccretion-disk}Accreted angular momentum $J_\mathrm{acc}$ and resulting rotational velocity $v_\mathrm{rot,MS}$ and velocity increase $\Delta v$ at $10^5\,\yr$ for our default model ($\mu_\mathrm{B}=2\times10^{37}\,\mathrm{G}\,\mathrm{cm}^3$) and the model with larger magnetic field ($\mu_\mathrm{B}=1\times10^{40}\,\mathrm{G}\,\mathrm{cm}^3$). A total mass of $\Delta M=0.1\,\msun$ is accreted when the merger models reach a luminosity of $\log L/\lsun=4.7$ and $5.1$. The stellar radii $R$ of the models at the time of accretion are also provided.}
\centering
\begin{tabular}{lccccc}
\toprule 
$\mu_\mathrm{B}$ & $\log L/\lsun$ & $R$ & $J_\mathrm{acc}$ & $v_\mathrm{rot,MS}$ & $\Delta v$ \\
$[\mathrm{G}\,\mathrm{cm}^3]$ &  & $[\rsun]$ & $[\mathrm{g}\,\mathrm{cm}^2\,\mathrm{s}^{-1}]$ & $[\kms]$ & $[\kms]$ \\
\midrule
\midrule 
$2\times10^{37}$ & 4.7 & 5.7 & $2.55\times10^{51}$ & 79 & 26 \\
$2\times10^{37}$ & 5.1 & 10.5 & $3.35\times10^{51}$ & 88 & 35 \\
$1\times10^{40}$ & 4.7 & 5.9 & $0.19\times10^{51}$ & 9 & 4 \\
$1\times10^{40}$ & 5.1 & 10.3 & $1.20\times10^{51}$ & 11 & 6 \\
\bottomrule
\end{tabular}
\end{table}

A summary of the accreted angular momentum and the resulting changes of the surface rotational velocities at $t=10^5\,\yr$ is provided in Table~\ref{tab:reaccretion-disk}. The time $t=10^5\,\yr$ roughly corresponds to the time when the merger models have reached the MS again. As expected, the accreted angular momentum and final rotational velocities are larger if the star accretes the mass when it is more expanded. Also, the accreted angular momentum is smaller and the final rotation is slower in the high magnetic-field cases because of the formation of a magnetosphere. In total, the rotational velocity of our default model will increase by ${\approx}\,50\%$ if $0.1\,\msun$ are accreted. In the more magnetic model, magnetic braking is efficient and the accreted angular momentum is lost again well before $10^6\,\yr$. The merger model with the weaker magnetic field essentially behaves like our default model upon mass accretion.

\section{Discussion}\label{sec:discussion}

\subsection{Spin of merger product}\label{sec:discussion-final-spin}

The final spin of the merger product depends on two aspects: first, how much angular momentum ends up in the merger product (Sect.~\ref{sec:discussion-initial-J}) and, second, how much angular momentum can be lost and gained during the thermal relaxation phase and the subsequent evolution (Sect.~\ref{sec:discussion-J-loss-gain}). The models presented in Sect.~\ref{sec:merger-evolution} cover various scenarios that show that the merger product might be a slow rotator but could also be a more rapid rotator after thermal relaxation.

\subsubsection{Initial angular momentum of merger product}\label{sec:discussion-initial-J}

As discussed in Sects.~\ref{sec:3d-to-1d} and~\ref{sec:spin-evolution}, the initial angular momentum of the merger product is set by the (viscous) co-evolution of the central star and torus right after the merger. At the moment, it is unclear how much angular momentum is transported out of the star and through the torus but models of this evolutionary phase in the context of double white-dwarf mergers suggest that a substantial fraction of the angular momentum may be transported outwards efficiently \citep[\eg][]{2012ApJ...748...35S,2012MNRAS.427..190S,2013ApJ...773..136J}. 

The initial rotation profile of the merger product is motivated by the idea that the strong magnetic fields establish solid-body rotation and that the surface of the merger product rotates near the Keplerian value because of locking to the torus (both is indeed realised in the \threed model after the merger). So far, we did not consider viscous heating and the corresponding entropy changes caused by the accretion of material from the torus. This could further heat the envelope of the merger and may lead to a larger radius. With a larger radius, a merger product with solid body rotation profile can accommodate more angular momentum because the total angular momentum in such a case scales with $\sqrt{R}$ (Eq.~\ref{eq:init-ang-mom}). A factor of 2 larger radius would increase the initial angular momentum by ${\approx}\,40\%$, \ie it would only increase from the current 3\% to maybe 4\%--5\% in terms of the total initially available angular momentum.

\subsubsection{Angular-momentum loss and gain}\label{sec:discussion-J-loss-gain}

We have identified several mechanisms by which the merger product can lose and gain angular momentum during the thermal relaxation and thereafter. All our models reach surface critical rotation such that they have to lose angular momentum. As shown in Sect.~\ref{sec:merger-evolution}, the amount of mass and angular-momentum loss in this phase depends on the timescale on which angular momentum is transported through the interior of the star to the surface. In our models, this timescale is set by the angular-momentum transport through the large-scale magnetic field and therefore depends on its strength. With the large magnetic dipole moment of $\mu_\mathrm{B}=1\times10^{40}\,\mathrm{G}\,\mathrm{cm}^3$, the model is able to shed nearly all its angular momentum already during the relaxation phase. This is even true if the initial angular momentum is increased to about 30\% of the total available angular momentum of the merger. These highly magnetic models further show that magnetic braking spins down the star already before reaching the MS (and of course also thereafter) such that the final merger product always rotates very slowly ($v_\mathrm{rot}<1\,\kms$).

Other studies of magnetic fields in stellar interiors \citep[\eg][]{2005ApJ...626..350H, 2011A&A...530A.115B, 2012MNRAS.424.2358P, 2015ApJ...799...85W, 2018MNRAS.477.2298Q} model the evolution of (usually) the poloidal and toroidal magnetic-field components, whereas we apply a fixed magnetic field. In our models, the magnetic fields lead to solid-body rotation and transport angular momentum to the surface, where it can then be lost by winds and magnetic braking. This is generally similar to the aforementioned studies: the effective net result regarding the final spin of the merged star in our model is expected to be similar to models with a more detailed treatment of the magnetic-field evolution. While this is true for the final spin of the star after the thermal relaxation, our model lacks predictive power for how the magnetic field evolves during the star's further life up to core collapse (see Sect.~\ref{sec:discussion-survival-b-fields}).

In our default model ($\mu_\mathrm{B}=2\times10^{37}\,\mathrm{G}\,\mathrm{cm}^3$), magnetic braking is negligible. However, if the field was stronger right after the merger and then dropped to the value observed today in \tausco, magnetic braking could be more important as is demonstrated in Fig.~\ref{fig:time-evol-mu-1e40} for our highly magnetic merger model. Furthermore, magnetic braking becomes more efficient if the merger product still rotates significantly after thermal relaxation. In this case, the star evolves on a nuclear timescale which can be comparable and even longer than the spin-down timescale due to magnetic braking. Models of initially rapidly rotating stars that incorporate magnetic braking have shown efficient spin-down \citep[\eg][]{2011A&A...525L..11M,2012MNRAS.424.2358P,2020MNRAS.tmp..227K} and such an evolution may be anticipated for our merger models that still rotate fast after thermal relaxation. 

In principle, a merger product could exceed the Eddington limit during the thermal relaxation phase because of energy deposition in the star from the coalescence. In our case, its luminosity never reaches the Eddington limit but other (more massive) mergers may do. At maximum luminosity, the merger model is in the S~Doradus instability strip of quiescent luminous blue variables \citep[\eg][]{2017RSPTA.37560268S}. An instability similar to that observed by \citet{2018Natur.561..498J} may occur but our merger model only briefly reaches the high luminosities of the S~Doradus instability strip. In conclusion, it is conceivable that merger products in general may experience phases with super-Eddington winds or eruptions during their thermal relaxation that could lead to significant mass and angular-momentum loss. The star \etacar might be an example of such an evolution through a super-Eddington phase caused by a merger \citep[\eg][]{2018MNRAS.480.1466S,2019MNRAS.485..988O,Hirai+2020}. 

Finally, some mass and angular momentum could be re-accreted from a left-over disk if the disk is not evaporated by the radiation of the merger product. In our default model, at most $0.1\,\msun$ may be re-accreted. Still, this can account for a non-negligible fraction of the angular momentum of the merger product (see Sect.~\ref{sec:reaccretion-disk}) because the mass could be accreted with a large specific angular momentum when the star is still bloated.

We conclude that there are several ways of removing and adding angular momentum, rendering clear predictions for the final spin of the merger product difficult at the moment. It is however intriguing (while somewhat counter-intuitive) that the merger product after thermal relaxation could well be a slow rotator.

\subsection{Surface enrichment with nuclear burning products}\label{sec:discussion-surface-enrichment}

Our models do not show a significant surface enrichment with nuclear burning products such as helium and nitrogen. However, enriched layers could be exposed in at least four ways. First, if more mass remains in the torus and is not accreted such that the surface of the central star is composed of layers more enriched with nuclear burning products. Second, by mixing of the torus while it is accreted onto the central star. Third, by mass loss during the thermal relaxation (when the star rotates critically or via winds and eruptions) and, fourth, by additional mixing after torus accretion. This could, \eg, be achieved if the large-scale magnetic fields not only contribute to angular-momentum transport but also to chemical mixing as argued by \citet{2002A&A...381..923S} for their dynamo operating in radiative regions of stars, or if the usual rotational mixing is efficient. 

For example, if the material accreted from the torus was fully mixed, the surface-nitrogen mass fraction of the merger product would be $0.00183$, \ie enhanced by a factor of ${\approx}\,2.5$ compared to the base nitrogen abundance of the models. Similarly, nitrogen at the surface would be enhanced by factors of ${\approx}\,3.2$ and ${\approx}\,2.0$ if only $1\,\msun$ and $2\,\msun$ of the torus are accreted, respectively (instead of the assumed $2.9\,\msun$). In light of these arguments, the non-enriched surfaces of our merger models are not a strong prediction and it is conceivable that the surface can be more enriched in nuclear burning products than our default models currently predict.

In such a case, our merger model would offer a possibility to explain the puzzling slowly-rotating and nitrogen-enriched stars in the Hunter diagram \citep{2008ApJ...676L..29H}. Other models trying to explain these objects invoke initially fast rotators with magnetic fields where the rotation mixes nitrogen to the surface and the magnetic fields subsequently spin down stars via magnetic braking \citep[\eg][]{2011A&A...525L..11M,2012MNRAS.424.2358P,2020MNRAS.tmp..227K}.

\subsection{Circumstellar nebulae and mass-ejection episodes}\label{sec:discussion-nebulae}

Before, during and after the actual merger there are several episodes where mass can be ejected to form a nebula. Before the merger when the binary is in deep contact, mass is lost through the outer Lagrangian points and can form an equatorial outflow \citep[\eg][]{2017ApJ...850...59P} that could form a circumbinary disk \citep{2018ApJ...868..136M}. In the merger itself, some mass is dynamically lost (in our case $<0.1\,\msun$ for tightly bound progenitors) and interacts with the matter lost during the previous spiral-in phase. In some merger cases, such matter has been shown to be deflected to produce bipolar nebulae \citep[\eg][]{2006MNRAS.365....2M,2007Sci...315.1103M,2009MNRAS.399..515M,2018ApJ...868..136M}. Indeed, bipolar nebulae are observed in past merger events \citep[\eg][]{2018NatAs...2..778K}. In this context, \citet{2018MNRAS.480.1466S} interpreted the bipolar Homunculus nebula of $\eta$~Car to be caused by a merger.

Later during the thermal relaxation phase of the merger, critical rotation is reached in our models and more massive mergers might also exceed the Eddington limit or experience envelope instabilities. This will result in further mass loss and maybe even the onset of a super-Eddington wind or eruptive phases \citep{2017MNRAS.472.3749O}. Rotationally-driven mass loss may operate predominantly in the equatorial plane but super-Eddington winds of rapidly-rotating stars could be even stronger at the poles than at the equator because of gravity darkening. Again, this could form bipolar nebulae. These winds could blow into previous ejecta and further shape the nebulae of mergers. Taken together, there could be several distinct mass-loss/ejection phases that may form nebulae.

In the context of mergers as progenitors of magnetic massive stars, a bipolar nebula surrounding a magnetic massive star may therefore be considered a smoking gun for the merger hypothesis. To our knowledge, no such nebula has been reported around \tausco but given that in our model the merger and nebula formation occurred $1\text{--}2\,\myr$ ago, it might be hard to still detect tracers of this. HD\,148937, a magnetic O6.5f?p star with a massive ($2\,\msun$), nitrogen enriched bipolar nebula \citep{1987A&A...175..208L,2010A&A...520A..59N}, might be such a smoking gun as suggested by \citet{2012ARA&A..50..107L}. The star is a rather fast rotator ($7\,\mathrm{d}$ rotation period) with a strong surface magnetic field \citep[$B_\mathrm{p}=100\text{--}300\,\mathrm{G}$;][]{2010A&A...520A..59N}. The age of the nebula has been estimated from its expansion velocity to be of order $3,000\,\yr$ \citep{1987A&A...175..208L} which would imply that the star is most likely in its thermal relaxation phase. Depending on in which phase the star is, it could spin-down further through internal density readjustments and magnetic braking (\cf Sect.~\ref{sec:merger-evolution}). There is a second nebula surrounding HD\,148937 (just as in the case of $\eta$~Car), which could be understood to be due to the several mass-ejection episodes discussed above.

\subsection{Survival of magnetic fields}\label{sec:discussion-survival-b-fields}

The magnetic field produced in the merger is thought to be long-lived \citep{2019Natur.574..211S}. Ohmic dissipation is slower than the lifetime of massive stars because of the high conductivity of stellar interiors. \citet{2004Natur.431..819B} and \citet{2006A&A...450.1077B} show that magnetic field structures with linked poloidal and toroidal components are stable and exist for about the main-sequence phase of their studied A-type star. They observe that these interlinked field components slowly rise to the stellar surface and may at some point disappear. Analytical approaches support these ideas \citep{2010MNRAS.405.1845M} and some stellar evolution computations with dynamo-generated magnetic fields also show a decrease of the surface field strength over time \citep[\eg][]{2012MNRAS.424.2358P, 2018MNRAS.477.2298Q}. Thus at some point in the evolution of a magnetic star, there might no longer be an observable, strong surface magnetic field, and such a star may then be considered as ``non-magnetic''. Such a process could be responsible for the apparent dearth of evolved magnetic stars (\eg \citealt{2007A&A...470..685L}, \citealt{2016A&A...592A..84F}, but see also \citealt{2019MNRAS.489.5669P}). Our models do not capture such a possible evolution because we assume a fixed magnetic field.

In the further evolution of our merged star towards core collapse, the star develops various convective regions driven by nuclear burning in and near the core and high opacity in the envelope. \cite{2004MNRAS.355L..13T} argue that convective regions expel or even destroy magnetic fields, but as long as there are radiative regions left in which the magnetic field can survive, it may prevail until core collapse. 

These ideas are supported by the detailed stellar-evolution computations of \citet{2018MNRAS.477.2298Q}. In their study, the authors follow the dynamo-generated, poloidal and toroidal magnetic-field components in a $3\,\msun$ star up to the end of the asymptotic-giant-branch phase. While the dynamo-generated fields are suppressed in convective regions, they are sustained in radiative regions and particularly in the degenerate core until the end of the computation. Because massive stars as those discussed here are never fully convective, there is reason to believe that at least parts of the magnetic field also exist at core collapse. 

In the late burning stages after core helium exhaustion, the evolution of stars speed up because of neutrino losses. Various convective regions driven by carbon, neon, oxygen and silicon burning show up on short timescales, and it remains to be seen how these convective regions interact with the magnetic field. For example, the magnetic field may be able to diffuse quickly enough from one radiative region to another before it can be captured, tangled and maybe even destroyed by newly formed convective regions. Also, a strong magnetic field may even survive inside convective regions, \eg in flux tubes, and leave the surrounding convective fluid motions almost unaffected \citep[see \eg][]{2017RSOS....460271B}.

\subsection{Final fate of merger product}\label{sec:discussion-final-fate}

Our merger model evolves closely to that of a genuine single star and will undergo core collapse as a red supergiant with a hydrogen-rich envelope. The ensuing supernova is therefore of Type~II. For a stellar population with a primordial binary fraction of 50\%, \citet{2019arXiv190706687Z} estimate that about 31\% (12\%--44\%) of all Type~II supernova stem from mergers and mergers of MS stars as discussed here may make up 6\% (1\%--17\%) of all Type~II supernovae. 

If the magnetic fields amplified in such mergers prevail up to core collapse (Sect.~\ref{sec:discussion-survival-b-fields}), they are expected to influence the supernova mechanism and the gas flow as is shown by the high magnetic-field case studied by \citet{2014MNRAS.445.3169O}, who find energy equipartition of the compressionally amplified magnetic fields and the gas flow. This leads to an earlier onset of the explosion because the strong magnetic field tends to guide the neutrino heated convective bubbles. The magnetic fields may suppress the standing accretion shock instability \citep[SASI;][]{2010ApJ...713.1219E} and thereby could even influence the kick that a proto-neutron star receives. However, whether the magnetic fields stabilise or destabilise the SASI appears to depend on the exact field geometry \citep[\eg][]{2010ApJ...711...99G,2014MNRAS.445.3169O}.

As shown by \citet{2019Natur.574..211S}, if the magnetic flux of $4\times10^{28}\,\gauss\,\mathrm{cm}^2$ within the innermost $1.5\,\msun$ right after the merger is preserved until core collapse, a neutron star with $10\,\mathrm{km}$ radius would have a magnetic field of about $10^{16}\,\gauss$ at the surface. So even if only 1\% of the magnetic flux is conserved up to core collapse, the field strength would still be of order $10^{14}\,\gauss$, \ie well within typical surface B-fields inferred for magnetars \citep[$10^{13}\text{--}10^{15}\,\gauss$,][]{2014ApJS..212....6O}. This makes our merger scenario a promising channel for the formation of magnetars. Furthermore, as shown by \citet{2019Natur.574..211S}, the incidence of mergers among massive stars and the formation rate of magnetars in supernovae appear to be similar.

\citet{2014MNRAS.437..675W} noted that the magnetic flux per unit mass, $\Phi_\mathrm{P}/M$, of magnetic OBA stars and highly-magnetic white dwarfs are rather similar and fall in the range $10^{-8.5} \text{--} 10^{-6.5}\,\gauss\,\mathrm{cm}^2\,\mathrm{g}^{-1}$. Because of the similarities of these magnetic fluxes, \citet{2014MNRAS.437..675W} suggest a common mechanism that may have generated the magnetic fields in these stars. From their \threed MHD merger simulation and under the assumption of magnetic-flux conservation, \citet{2019Natur.574..211S} predict a surface magnetic-field strength of ${\approx}\,9\,\mathrm{kG}$ for the $16.9\,\msun$ merger product with a radius of $5\,\rsun$ on the MS, and ${\approx}\,10^{16}\,\mathrm{G}$ for a $1.5\,\msun$ neutron star of radius $10\,\mathrm{km}$. Computing the magnetic flux per unit mass as defined by \citet{2014MNRAS.437..675W}, \ie $\Phi_\mathrm{P}/M \equiv R^2 B_\mathrm{P}/M$ with the stellar radius $R$, the poloidal magnetic field $B_\mathrm{P}$ and the mass $M$, we find $\Phi_\mathrm{P}/M\,{\approx}\,3\times10^{-8}\,\gauss\,\mathrm{cm}^2\,\mathrm{g}^{-1}$ and $\Phi_\mathrm{P}/M\,{\approx}\,4\times10^{-6}\,\gauss\,\mathrm{cm}^2\,\mathrm{g}^{-1}$ for the $16.9\,\msun$ MS merger product and the $1.5\,\msun$ neutron-star remnant, respectively. These values tend to agree with the above typical $\Phi_\mathrm{P}/M$ values of magnetic OBA stars and highly-magnetic white dwarfs. Hence, if the similarity of the magnetic fluxes indeed point to a common origin of the magnetic fields in MS OBA stars, white dwarfs and magnetars, the MS merger model studied here and in \citet{2019Natur.574..211S} would fit into this picture and could constitute one such common origin.

Magnetar-powered explosions have been suggested to be responsible for some long-duration gamma-ray bursts and superluminous supernovae \citep[\eg][]{2009MNRAS.396.2038B,2010ApJ...717..245K}. However, the highly-magnetised cores of such engine-driven models are unlikely to stem from MS mergers, because magnetar formation from our models is too frequent: from the incidence of magnetic massive stars we expect a magnetar formation rate of about 10\% of the core-collapse supernova rate \citep{2019Natur.574..211S} while long-duration gamma-ray bursts and superluminous supernovae occur only with a rate of 0.01\%--0.1\% of that of core-collapse supernovae \citep[\eg][]{2018SSRv..214...59M}. Also, our default model is a slow rotator at core collapse and therefore does not provide the required rotational energy that seems to be needed for such rare and energetic transients \citep[see \eg][]{1993ApJ...405..273W}. However, if a merger remnant with a highly magnetised core accretes angular momentum in a late evolutionary stage (\eg from an original tertiary star), the merger remnant could be rapidly rotating and highly magnetised at core collapse. Alternatively, if the cores of two stars in a common-envelope merge, the merger may produce strong magnetic fields and angular momentum may be retained as suggested by \citet{2011MNRAS.410.2458T}. Such scenarios are probably much rarer and thus could be viable for progenitors of long-duration gamma-ray bursts and some superluminous supernovae.

\section{Conclusions}\label{sec:conclusions}

We study the long-term evolution of the product of a \threed MHD merger simulation of two massive main-sequence stars with the \oned stellar evolution code \mesa. Strong magnetic fields have been produced in the merger and we consider the influence of these magnetic fields on the evolution of the merger product through interior angular-momentum transport and additional angular-momentum loss from the surface (magnetic braking).

Directly after the merger, the star is out of thermal equilibrium which causes an initial expansion and then contraction phase that finally brings the star back to the main sequence. During this thermal relaxation, which lasts for $10^4\text{--}10^5\,\yr$, the merger product reaches break-up rotation at the surface while expanding. The surface spin-up during the expansion phase is caused by internal angular-momentum transport because of the magnetic field and restructuring of the density profile. Somewhat counter-intuitively, the star spins down while contracting back to the main sequence. The spin-down is mainly driven by changes in the moment of inertia factor. In our models with rather modest magnetic fields, magnetic braking is unimportant for the evolution but becomes crucial for stronger fields.

While approaching thermal equilibrium, the star drives an unusually large convective core which mixes further fresh hydrogen fuel into the core and thereby contributes significantly to its overall rejuvenation. Once in full equilibrium, the merger model is a slow rotator and it evolves similar to a genuine single star of the same mass. The envelope of the merger product is enriched with helium and other hydrogen-burning products which makes it over-luminous compared to a single star of the same mass. Despite the helium-rich envelope, no nuclear-processed material is found at the surface but this is very model dependent and it is conceivable that nuclear-burning products such as nitrogen appear on the surface. Once back on the main sequence, the merger model is compatible with the magnetic and rejuvenated star \tausco and therefore offers a promising way to explain its magnetic field and general appearance as a rejuvenated, massive blue straggler in the Upper Scorpius association.

The spin evolution and surface chemical abundances of the merger product are still prone to considerable uncertainties. While magnetic braking and other angular momentum loss and gain processes are important, the spin of the merger product on the main sequence is probably mainly set during the viscous accretion phase of the torus left behind by the merging process.

Altogether, we have shown that it is feasible that some merger products are slow rotators, contrarily to usual expectations. If these magnetic stars can keep a magnetised core during their evolution up to core collapse, they may leave magnetar remnants after their final supernova explosion. Analogously, we expect that lower-mass, magnetised merger products could form highly-magnetic white dwarfs at the end of their lives.

\section*{Acknowledgements}

We thank the anonymous reviewer for helpful comments. 
This work was supported by the Oxford Hintze Centre for Astrophysical Surveys which is funded through generous support from the Hintze Family Charitable Foundation. 
STO, FKR and FRNS acknowledge funding from the Klaus Tschira foundation.


\bibliographystyle{mnras}
\newcommand{\noop}[1]{}



\appendix

\section{Evolution with different magnetic field strengths}\label{sec:evol-different-b-fields}

Our default model assumes a magnetic dipole moment of $\mu_\mathrm{B}=2\times10^{37}\,\mathrm{G}\,\mathrm{cm}^3$ (see Sect.~\ref{sec:b-field}). The evolution of the merger product for a three order of magnitude lower dipole moment ($\mu_\mathrm{B}=2\times10^{34}\,\mathrm{G}\,\mathrm{cm}^3$) in the HR diagram and key properties of the model are shown in Fig.~\ref{fig:hrd-mu-2e34} and~\ref{fig:time-evol-mu-2e34}, respectively. As for the default model, the magnetic field mainly influences the interior angular-momentum transport and magnetic braking is insignificant. With the weaker magnetic field, the spin-down occurs later but the model after thermal relaxation is also a slow rotator that appears to be compatible with observations of \tausco. 

For a magnetic dipole moment stronger by about three orders of magnitude than the default model ($\mu_\mathrm{B}=1\times10^{40}\,\mathrm{G}\,\mathrm{cm}^3$), there are qualitative differences (see Figs.~\ref{fig:hrd-mu-1e40} and~\ref{fig:time-evol-mu-1e40}): magnetic braking is relevant now such that the final spin of the merger product after thermal relaxation is even slower. The chosen large dipole moment is reminiscent of the magnetic field strength observed at the end of the \threed MHD simulations \citep{2019Natur.574..211S} but the model on the MS would have an unrealistically strong magnetic field that even exceeds that of $34\,\mathrm{kG}$ of the current record holder, Babcock's star HD~215441 \citep{1960ApJ...132..521B}.

\begin{figure}
\begin{centering}
\includegraphics[width=0.49\textwidth]{{{hrd-M-16.90-mu-2e34-no-gradT-excess}}}
\par\end{centering}
\caption{As Fig.~\ref{fig:hrd-mu-2e37} but for a magnetic dipole moment of $\mu_\mathrm{B}=2\times10^{34}\,\mathrm{G}\,\mathrm{cm}^3$.}
\label{fig:hrd-mu-2e34}
\end{figure}

\begin{figure}
\begin{centering}
\includegraphics[width=0.49\textwidth]{{{hrd-M-16.90-mu-1e40-no-gradT-excess}}}
\par\end{centering}
\caption{As Fig.~\ref{fig:hrd-mu-2e37} but for a assumed magnetic dipole moment of $\mu_\mathrm{B}=1\times10^{40}\,\mathrm{G}\,\mathrm{cm}^3$.}
\label{fig:hrd-mu-1e40}
\end{figure}

\begin{figure}
\begin{centering}
\includegraphics[width=0.49\textwidth]{{{time-evol-M-16.90-mu-2e34-no-gradT-excess}}}
\par\end{centering}
\caption{As Fig.~\ref{fig:time-evol-mu-2e37} but for an assumed magnetic dipole moment of $\mu_\mathrm{B}=2\times10^{34}\,\mathrm{G}\,\mathrm{cm}^3$.}
\label{fig:time-evol-mu-2e34}
\end{figure}

\begin{figure}
\begin{centering}
\includegraphics[width=0.49\textwidth]{{{time-evol-M-16.90-mu-1e40-no-gradT-excess}}}
\par\end{centering}
\caption{As Fig.~\ref{fig:time-evol-mu-2e37} but for an assumed magnetic dipole moment of $\mu_\mathrm{B}=1\times10^{40}\,\mathrm{G}\,\mathrm{cm}^3$.}
\label{fig:time-evol-mu-1e40}
\end{figure}

\bsp	
\label{lastpage}
\end{document}